\documentclass[aps,pra,twocolumn,10pt]{revtex4-2}

\usepackage{graphicx}
\usepackage{dcolumn}
\usepackage{bm}
\usepackage{amsmath, amssymb}
\usepackage{tikz}
\usepackage{etoolbox}
\apptocmd{\sloppy}{\hbadness 10000\relax}{}{}
\usepackage{pgfplots} 
\pgfplotsset{compat=1.16}

\newcommand{\ket}[1]{\lvert#1\rangle}
\newcommand{\bra}[1]{\langle#1\rvert}
\newcommand{\sca}[2]{\langle#1\vert#2\rangle}
\newcommand{\abs}[1]{\left\lvert #1\right\rvert}

\begin{document}

\preprint{APS/123-QED}

\title{Quantized and nonquantized Hall response in topological Hatsugai-Kohmoto systems}

\author{Thibaut Desort}
\affiliation{
 Laboratoire de Physique des Solides, Universit\'e Paris-Saclay, CNRS UMR 8502, F-91405 Orsay cedex, France
}

\author{Mark O. Goerbig}
\affiliation{
 Laboratoire de Physique des Solides, Universit\'e Paris-Saclay, CNRS UMR 8502, F-91405 Orsay cedex, France
}

\author{Corentin Morice}
\affiliation{
 Laboratoire de Physique des Solides, Universit\'e Paris-Saclay, CNRS UMR 8502, F-91405 Orsay cedex, France
}

\date{\today}

\begin{abstract}
We explore the robustness of Hall conductivity quantization in several insulating systems, exhibiting one scenario where the quantization is not preserved. Specifically, we apply the Kubo formula to topological models with the Hatsugai-Kohmoto interaction. Starting from the many-body degeneracy induced by this interaction in the topological Kane-Mele model, we consider Zeeman fields to select specific states within the ground-state manifold that reveal a non-quantized Hall response, precisely for the case with a Zeeman field diagonal in the bands of the Kane-Mele model. From a physical point of view, this term may mimic a ferromagnetic order that arises naturally when couplings beyond the Hatsugai-Kohmoto interaction are taken into account.
\end{abstract}

\maketitle

\section{Introduction}

Single-particle topological band theory \cite{Cayssol2021} links the Hall conductivity $\sigma_{xy}$ of topological insulators to the Chern number $\mathcal{C}$, a $\mathbf{Z}$-topological invariant,  $\sigma_{xy} =\mathcal{C}e^2/h$. This has provided a successful explanation of the integer quantum Hall effect, where the transverse conductivity for materials in a strong magnetic field is quantized \cite{Thouless1982}, or else of the quantum anomalous Hall effect in the absence of a magnetic field \cite{Haldane1988}. In spite of its success, topological band theory cannot, strictly speaking, cover many experimentally relevant situations where the properties of electrons in quantum matter are strongly influenced or even driven by electronic interactions \cite{Klitzing1980}. To account for any quantization of the Hall conductivity for interacting systems, the original theoretical description needs to be extended to take into account the yet poorly understood articulation between electronic correlations and band topology.

A seminal paper by Niu, Thouless and Wu  (NTW) \cite{Niu1985} generalizes, using the Kubo formula, the quantization of the Hall conductivity in the presence of local interactions, such as the Coulomb interaction. This is valid both when the ground state is unique and when the ground-state manifold exhibits topological, non-extensive degeneracy, for instance in the case of topological order. The new $\mathbf{Z}$-topological invariant is no longer the band Chern number, but rather the many-body generalization of it, expressed as an integral over twisted boundary conditions \cite{Kudo2019,Goldman2024}. In the non-interacting case, however, it reduces to the Chern number.
 
To see if physical observables in many-body systems are quantized or not, one can express them as simple functions of topological invariants. To compute these invariants, topological field theory \cite{Volovik2003,Wang2010,Wang2012} employs Green's functions because generally, for interacting systems, the eigenstates cannot be computed. However, some of these invariants do not always coincide with the Hall conductivity or another well identified measurable response function \cite{PeraltaGavensky2023,Zhao2023,Yang2019,Bollmann2024}. There is a growing body of evidence that the breakdown of identities between observables and topological invariants is directly connected to the presence of bands of zeros of the Green's function \cite{PeraltaGavensky2023,Zhao2023,Yang2019,Bollmann2024}. Indeed, it has since long been known that Green's functions encode topological information in their poles \cite{Gurarie2011} but the study of their zeros has become increasingly systematic \cite{Wagner2023,Blason2023,Setty2024,Simon2025}. Additionally, the formulas for topological invariants involve products of multiple Green's functions, making them difficult to apply in practice.
Because of this, generic tools have been developed to extract topological information more easily: for non-degenerate systems, the zero-frequency Green's function, also known as the topological Hamiltonian \cite{Wang2013}, can be interpreted, because it is Hermitian, allowing one to use the same theoretical tools as in non-interacting systems. This demonstrates that, under certain conditions, the Hall conductivity remains quantized even in the presence of interactions.
  
These findings highlight the growing interest in the interplay between strong correlations and topology \cite{Hohenadler2013,Rachel2018}. In this context, systems including the Hatsugai-Kohmoto (HK) interaction \cite{Hatsugai1992,Zhao2025}, whose Green’s functions can be computed exactly \cite{Manning-Coe2023}, have received significant attention. This extended even to the field of cuprates, since HK systems mimic several important features of the pseudogap phase \cite{Worm2024}.
The spinful Haldane model \cite{Haldane1988}, combined with this HK interaction, exhibits a coexistence of a Mott phase and a topological insulating phase, with the associated quantized Hall conductivity \cite{Mai2023a,Mai2023}. This topological Mott insulating state has also been reported in systems with HK interaction diagonal in orbitals \cite{Jablonowski2023}. More generally, the connection between topological invariants, edge states, strongly correlated phases of matter and Hall conductivity is now widely studied in HK models \cite{Wysokinski2023,Mai2024,Skolimowski2025}.

The aim of the present paper is to investigate under which circumstances topological band models with a HK interaction exhibit or not quantized Hall response, in the form of a quantized Chern number. For this purpose, we use the Kane-Mele model \cite{Kane2005} and a HK interaction that is diagonal in the non-interacting bands in order to perturb their original topological character. We concentrate on two-band quarter-filled systems, i.e., where the valence band hosts one electron per wave-vector state. We lift the macroscopic spin degeneracy using the Zeeman effect, which we investigate in two different variants, one being diagonal in the (local) orbital basis and the other one in the band basis. Our main finding is a robust quantization of the Hall response of topological HK models in the case of an orbital Zeeman term, while the band Zeeman term yields a response that varies continuously between $-e^2/h$ and $+e^2/h$.

The paper is organized as follows. In Sec.~\ref{sec:Hall}, we review essential results on the HK interaction and the computation of the Hall conductivity via the Kubo formula. We particularly focus on how the HK interaction generates a degeneracy that can be used to control the Hall conductivity (Sec.~\ref{sec:Hallbis}). The HK interaction, in this case, serves as a platform to test the effects of many-body degeneracy on this observable. We then discuss the results for non-interacting topological spinful Haldane and Kane-Mele models, in Sec.~\ref{sec:TM}. In Sec.~\ref{sec:zeeman}, starting from the Kane-Mele model, we introduce a Zeeman field in two forms: diagonal in the atomic orbitals in one case and diagonal in the bands in the other. This Zeeman term selects an orientation for the spins of the system. While the former arises directly in physically relevant situations, the latter can represent a natural ferromagnetic order on the mean-field level if couplings beyond the pure HK interaction are taken into account perturbatively. We analyze how this affects the behaviour of the Berry curvature and the Hall conductivity. Finally, we discuss the conditions that lead to a non-quantization of this transverse observable (Sec.~\ref{sec:conc}).

%%%%%%%%%%%%
\section{Hall conductivity}\label{sec:Hall}

In this first section, we recall some aspects of the Hall conductivity and its particularities in interacting systems. 

\subsection{General tools}
\label{sec:Hallbis}

\subsubsection{Hatsugai-Kohmoto models}

 Let $H_0$ be a non-interacting Hamiltonian \begin{equation}
	H_0 = \sum_{\mathbf{k}\lambda\sigma} \epsilon_{\mathbf{k}\lambda\sigma} n_{\mathbf{k}\lambda\sigma},
\end{equation} 
with the energy bands $\epsilon_{\mathbf{k}\lambda\sigma}$ given in terms of the lattice momentum $\mathbf{k}$, the band index  $\lambda$, and $\sigma\in\lbrace\uparrow,\downarrow\rbrace$ representing the physical spin. For multiple bands, the $\mathbf{k}$-diagonal HK interaction can be generalized in various ways. Here, we choose a version that is immediately diagonal in the band index \begin{equation}
	H_\text{HK} = U \sum_{\mathbf{k}\lambda} n_{\mathbf{k}\lambda\uparrow} n_{\mathbf{k}\lambda\downarrow},
\end{equation}  
where the constant $U$ sets the interaction energy scale. It is a Hubbard-type interaction, but instead of being local in direct space, it is ``local'' in reciprocal space. Because it is diagonal in the original bands, it preserves the eigenstates of the non-interacting Hamiltonian. When an electron with spin $\sigma$ is created at momentum $\mathbf{k}$ in band $\lambda$ (we have the one-particle state $c_{\mathbf{k}\lambda\downarrow}^\dagger\ket\varnothing$ of energy $\epsilon_{\mathbf{k}\lambda\downarrow}$), creating the partner electron at the same $\mathbf{k}$, that is, applying $c_{\mathbf{k}\lambda\uparrow}^\dagger$, costs $U+\epsilon_{\mathbf{k}\lambda\uparrow}$ rather than just $\epsilon_{\mathbf{k}\lambda\uparrow}$. The many-body eigenstates of the total Hamiltonian $H_0+H_\text{HK}$ remain Slater determinants, but they are intrinsically of many-body nature: to add an electron, you have to check if there is already an electron in the same orbital state with opposite spin. 

In contrast to the Hubbard interaction, which is local in real space and thus of stronger physical relevance than the HK interaction, the latter is of infinite range and thus yields correlations at arbitrarily large distances: it creates electron-hole pairs that are potentially separated by macroscopic distances, possibly at the edges of the system, as shown by Ref. \cite{Skolimowski2024}. Its main use is therefore to illustrate rather than to realistically simulate the role of electronic correlations. It is in this sense that we discuss it in the present paper, where we choose an analytical approach to discuss topological properties of correlated systems. 

More precisely, we study insulating two-band models at quarter-filling with a large $U$ (at least greater than the bandwidth of the lower band). At zero temperature, double occupancy is forbidden for each $\mathbf{k}$, meaning that there is exactly one particle per $\mathbf{k}$-sector. This is then a Mott insulator in reciprocal space \cite{Mai2023}: each particle is stuck at a given $\mathbf{k}$. Additionally, we focus on systems whose non-interacting part has spin-independent energies, i.e., $\epsilon_{\mathbf{k}\pm\uparrow}=\epsilon_{\mathbf{k}\pm\downarrow}\equiv\epsilon_{\mathbf{k}\pm}$. Consequently, at each $\mathbf{k}$, we have total freedom for the choice of the spin state because $c_{\mathbf{k}-\uparrow}^\dagger \ket{\varnothing}$ and $c_{\mathbf{k}-\downarrow}^\dagger \ket{\varnothing}$ are degenerate in energy. When we fill the system, any spin combination is valid thanks to this degeneracy.

The idea of the HK interaction is to populate only the lowest energy band in the spin-degenerate case. In section \ref{sec:zeeman}, we lift this degeneracy in the non-interacting Hamiltonian. The two initial degenerate bands have now different, but close, energies. However, they are not necessarily diagonal in spin so that, to only populate the lowest band, we have, following the original HK interaction, to forbid the occupancy of the second band when an electron occupies the lowest one. Labelling the increasing energies of the non-interacting Hamiltonian by $\lambda_i$  we define the interaction we use like \begin{equation}
	H_\text{HK} = U\sum_\mathbf{k} \left(n_{\mathbf{k}\lambda_1}n_{\mathbf{k}\lambda_2} + n_{\mathbf{k}\lambda_3}n_{\mathbf{k}\lambda_4}\right).
\end{equation} At quarter-filling, each $\mathbf{k}$ is populated with one electron in the lowest band, this is a projector on this band, exactly like the original HK interaction.

\subsubsection{Kubo formalism}

Under the hypothesis that the Green's functions vanish at large distances, NTW showed that the Hall conductivity is quantized \cite{Niu1985}. When the ground state is unique, it is quantized at an integer value, such as in the integer quantum Hall effect, whereas in topological order, it is quantized at a rational value, similarly to the fractional quantum Hall effect. This second case is more restrictive than a simple degeneracy: we need the operator connecting two ground states to be non-local, in the sense that it cannot involve a finite number of single-particle excitations. When the ground space does not fulfill one of these two conditions, the Hall conductivity is not necessarily quantized. 

Using the Kubo formula to compute the Hall conductivity results in \begin{equation}\label{eq:kubo}
	\sigma_{xy} = \frac{i e^2 \hbar}{V}\sum_{\nu=1}^d \sum_{n>0} \frac{\bra{0_\nu}v^x\ket{n}\bra{n} v^y\ket{0_\nu} - \text{c.c.}}{(E_n-E_0)^2},
\end{equation} where $V$ is the volume of the system and $d$ is the degeneracy of the ground space, $\lbrace\ket{0_\nu},\nu\rbrace$ are the corresponding ground states, indexed by $\nu$, and $E_0$ is their energy, $\ket{n}$ are the excited states and $E_n$ are their corresponding energies, $v^{x,y}$ are the components of the velocity operator.
The sum over the ground states highlights that changing the degeneracy of a system results in changing the Hall conductivity.

 To use the HK interaction for Hall conductivity computations, we have to define the velocity operator in the Kubo formula. Let us examine the (many-body) velocity operator defined by, for translationally invariant Hamiltonians, \begin{equation}\label{eq:def_vk}
	\mathbf{v}(\mathbf{k}) = \frac1{i\hbar} \left[H(\mathbf{k}),\mathbf{r}\right]=\frac1{i\hbar} \left[H_0(\mathbf{k}),\mathbf{r}\right]+\frac1{i\hbar} \left[H_\text{I}(\mathbf{k}),\mathbf{r}\right],
\end{equation} 
with the full Bloch Hamiltonian $H(\mathbf{k})$, including the interactions $H_\text{I}$, and the position operator $\mathbf{r}$. All physically relevant interactions are sufficiently local so that they commute with $\mathbf{r}$ hence one does not need to care about the last term of Eq.~(\ref{eq:def_vk}). The velocity operator and the current densities are therefore not affected by the last term and thus the interaction, the only role of which is to single out a ground state (or a set of ground states). Strictly speaking, this is not the case for the HK interaction, which is of infinite range so that corrections due to the last term should in principle be taken into account.  This turns out to be a subtle issue that has been the origin of a recent controversy about the conductivity and its topological origin in HK systems: Ref.~\cite{Guerci2025} explains that HK models suffer from a non-analycity of the current observable so that one cannot compute the Hall conductivity in such models. Ref.~\cite{Ma2025}, however, claims solving the problem by inverting two limits for the orbital HK model. However, in our situation, we study the band HK model which mixes the orbital degrees of freedom so that it is unclear that such result also applies here.

In the present work, we adopt a pragmatic or agnostic point of view where we consider the HK interaction only as a ``ground-state selector''\footnote{We are indepted to Alessandro Toschi who coined this term during common discussions.} and neglect possibly arising correcting terms to the velocity operator, which we restrict to the first term in Eq.~(\ref{eq:def_vk}),
\begin{equation}\label{eq:vk}
	\mathbf{v}(\mathbf{k}) =\frac1{i\hbar} \left[H_0(\mathbf{k}),\mathbf{r}\right] = \boldsymbol{\nabla}_\mathbf{k} H_0(\mathbf{k}).
\end{equation}  
In other words, in our approach, the HK interaction is only used for filling purpose: it does not impact the dynamics of the electrons, it only determines which states are filled by electrons. The non-locality of the HK interaction only enters through the many-body eigenstates of the system.

Moreover, the NTW result for the quantization of Hall conductivity has \textit{a priori} no reason to be true in HK systems, because the correlations do not vanish at large distances, as the interaction couples states that are arbitrarily distant, which was a key assumption for the NTW conclusion to be valid. Therefore, the systems we study are beyond the regime of validity of the theorem by NTW.

\subsubsection{Ground-state manifold}

At any value of $\mathbf{k}$, the two states $c_{\mathbf{k}-\uparrow}^\dagger \ket{\varnothing}$ and $c_{\mathbf{k}-\downarrow}^\dagger \ket{\varnothing}$ are degenerate. If we concentrate on the sector where all quantum states in the valence band $\lambda=-$ are singly occupied, the set of $N$-particle states 
\begin{equation}
\label{eq:canonical}
\lbrace\sigma_\mathbf{k} \rbrace=\left\lbrace\left(\prod_\mathbf{k} c_{\mathbf{k},\lambda=-,\sigma_\mathbf{k}}^\dagger\right) \ket\varnothing\right\rbrace,\end{equation}
with $\sigma_\mathbf{k}\in\lbrace\uparrow,\downarrow\rbrace$, forms a $2^N$-fold degenerate manifold. However, this situation is highly artificial and again due to the infinite-range character of the HK interaction: states at different, even infinitesimally close, wave vectors $\mathbf{k}$ and $\mathbf{k}+\delta\mathbf{k}$ are completely decoupled. 

This is an artificial situation, as highlighted by the results in Ref.~\cite{Mai2023a}, where it has been found that the magnetic susceptibility diverges at zero temperature, suggesting that only ferromagnetic states should be considered, instead of the whole manifold seen in Eq.~(\ref{eq:canonical}). This exact same calculation works for any system whose non-interacting part is spin-degenerate. The same result for strong local Hubbard interaction appears in Ref.~\cite{Tzeng2023} for the topological Bernevig-Hughes-Zhang model. The ferromagnetic subspace is the eigenspace of the total spin operator squared $\mathbf{S}^2$ with maximal eigenvalue, that is a vector space of dimension $(N+1)$ instead of the original dimension $2^N$. Among these states, we can choose those which are tensor-products in $\mathbf{k}$-space: these are states where all spins point in the same direction and they generate all the ferromagnetic states. However, to use the Kubo formula, we would need a basis for the orthogonal (excited) space. The problem is to generalize spin additions to a number $N$ of spins, which is not an easy problem. To avoid this complication, in the following, we keep this intuition for a ferromagnetic ordering and we mimic it (in a mean-field picture) by an external Zeeman field that equally singles out maximally polarized states and that allows us to use the Kubo formula in a simple manner.

\subsection{Hall conductivities in topological models}
\label{sec:TM}

We now investigate the HK interaction in two paradigmatic models whose non-interacting parts exhibit a topological behaviour, i.e., the Haldane model (including the spin degree of freedom) and the Kane-Mele model.

\subsubsection{Spinful Haldane model}

\begin{figure}[!ht]
	\centering
	\includegraphics[scale=1]{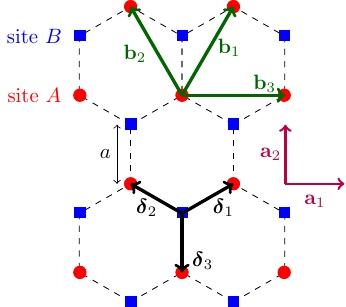}
	\caption{Structure of the honeycomb lattice and definition of the vectors $\boldsymbol{\delta}_i$.}
	\label{fig:structureGraphene}
\end{figure}

We recall the Haldane model \cite{Haldane1988} by parametrizing the honeycomb lattice as in Fig.~\ref{fig:structureGraphene}. 
We note $\Delta$ the Semenoff mass, $t$ the hopping to nearest neighbour and $t'$ and $\varphi$ the modulus and the argument of the hopping to second nearest neighbour. We consider the following Bloch Hamiltonian in the sublattice basis $\lbrace \ket{\mathbf{k}A},\ket{\mathbf{kB}}\rbrace$ \begin{equation}
	H(\mathbf{k}) = h_0(\mathbf{k})\mathbf{1}+\mathbf{h}(\mathbf{k})\cdot\boldsymbol{\sigma},\end{equation}
	with the different components, knowing $\gamma_\mathbf{k} = \text{e}^{\text{i}\mathbf{k}\cdot\boldsymbol{\delta}_1} + \text{e}^{\text{i}\mathbf{k}\cdot\boldsymbol{\delta}_2} + \text{e}^{\text{i}\mathbf{k}\cdot\boldsymbol{\delta}_3}$, \begin{equation} \quad\begin{cases}h_0(\mathbf{k}) = 2t'\sum_j \cos(\mathbf{k}\cdot\mathbf{b}_j)\cos(\varphi) \\ h_x(\mathbf{k}) = -t\mathfrak{Re}(\gamma_\mathbf{k}) \\ h_y(\mathbf{k}) = -t\mathfrak{Im}(\gamma_\mathbf{k}) \\ h_z(\mathbf{k}) = \Delta + 2t'\sum_j \sin(\mathbf{k}\cdot\mathbf{b}_j)\sin(\varphi)\end{cases}.
\end{equation} The eigenenergies of the model are then $h_0(\mathbf{k})+\epsilon_\mathbf{k}$ with $\epsilon_\mathbf{k} = \sqrt{\abs{t\gamma_\mathbf{k}}^2+h_z(\mathbf{k})^2}$. \\ We consider two copies of the Haldane model, one for spin $\uparrow$, one for spin $\downarrow$. The two key parameters for topology are the Semenoff mass $\Delta$ and the phase associated with the complex second-nearest-neighbour hopping $\varphi$, the variations of which define a topological phase diagram. We assume that the parameters are set such that we are in the topological phase where the Chern number is $\mathcal{C}=-1$. We retain the macroscopic degeneracy introduced by the HK interaction in this double Haldane model and apply the Kubo formula to the canonical basis defined in Eq.~(\ref{eq:canonical}).

In the Kubo formula, we have to compute matrix elements of the velocity operator. For any $\mathbf{k}$, this operator commutes with the number of particles $n_\mathbf{k}$, so that it cannot couple two states with a different number of particles. This immediately also excludes states with any double occupancy, because the ground states have one electron per $\mathbf{k}$-sector. We focus on one term of the sum, considering a given ground state $\ket{0_\nu}$ and an excited state $\ket{n}$. If $\ket{n}$ has a different spin state than $\ket{0_\nu}$, the term does not contribute because the velocity operator cannot flip spins. If $\ket{n}$ and $\ket{0_\nu}$ have the same spin state, the contribution is non-zero only if $\ket{n}$ is obtained from $\ket{0_\nu}$ by transferring an electron from momentum $\mathbf{k}$ in band $-$ to same momentum in band $+$. The calculation then proceeds similarly to the non-interacting case, yielding the quantized result, already obtained in Ref.~\cite{Mai2023a}, 
\begin{equation}
	\sigma_{xy}^\text{spinful Haldane} = -\frac{e^2}{h},\quad \mathcal{C}=-1.
\end{equation} 
In this case, quantization persists even in the presence of macroscopic degeneracy. This is indeed expected since the two spin branches of the valence band $\lambda=-1$ have the same geometric properties and thus the same Chern number. The latter is therefore unchanged if we change the spin orientation in one or several wave-vector states.

\subsubsection{Kane-Mele model}

We now consider the Kane-Mele model, which is constructed from two copies of the Haldane model such that $H_0^\downarrow(\mathbf{k})=H_0^\uparrow(-\mathbf{k})^*$, which enforces time-reversal symmetry. With this setup, using our notations, the terms $h_0$, $h_x$ and $h_y$ remain unchanged, but we now have \begin{equation}\label{eq:chiral}h_z(\mathbf{k},\downarrow) = h_z(-\mathbf{k},\uparrow).\end{equation} This differs from the spinful Haldane model because the spins $\uparrow$ and $\downarrow$ no longer play the same role. Therefore, one may have a Chern number of $-1$ for one spin branch of the valence band in the topological sector, while the other necessarily has a Chern number of $+1$ in order to respect time-reversal symmetry. 

We focus on the spin-degenerate case $\epsilon_{\mathbf{k}\pm\uparrow} =\epsilon_{\mathbf{k}\pm\downarrow}$ by setting $\Delta=0$. On this system, with the HK interaction, we calculate the Hall conductivity using the canonical basis and the same rules for excluding contributions we used in the previous paragraph. The result is also quantized but trivially zero due to time-reversal symmetry,
\begin{equation}
	\sigma_{xy}^\text{Kane-Mele} = 0.
\end{equation} In this model, electrons bearing spin $\uparrow$ move in one direction, while electrons bearing spin $\downarrow$ move in the opposite direction: they automatically have an opposite contribution as charge carriers. Because the two types of spin are equally mixed, as it is always the case in the Kane-Mele model, there is no net current.

\smallskip

%%%%%%%%%%%%%
\section{Macroscopic degeneracy breaking} \label{sec:zeeman}

In the following, we focus on the Kane-Mele model. We investigate the breaking of the macroscopic spin degeneracy by selecting ground states in order to change the average in the Kubo formula, thereby controlling the Hall conductivity. To achieve this, we introduce a Zeeman field, which preferentially selects one spin species over the other, for example going to ferromagnetic states. We consider the ferromagnetic phenomenology, coming from the divergence of the magnetic susceptibility, but this time generated by an external field that polarizes ferromagnetic states. This field may represent a magnetic order parameter in the ordered phase. We focus on the non-interacting parts of the model and, in the end, establish a connection with the Hall conductivity. 

We do not consider the spinful Haldane model anymore because the model is exactly identical for the two spin species. Introducing a Zeeman term selects a direction for the spin and this results in a mix of properties between spins up and down. Because the Hall conductivities are equal to $-1$ for both species, any mix of them will also be $-1$.

\subsection{Orbital Zeeman term}

First, we introduce a Zeeman field along a given axis in real space, which is diagonal in the orbital index $\alpha\in\lbrace A,B\rbrace$. After performing a Fourier transform, the corresponding Hamiltonian takes the form, noting the Zeeman field $\mathbf{B}$, \begin{equation}
	H_\text{orbital Zeeman} = -\mathbf{B}\cdot\sum_{\mathbf{k}\alpha} \mathbf{S}_{\mathbf{k}\alpha},\end{equation}
	where the spin operators are defined in terms of the Pauli matrices  $\boldsymbol{\sigma} = (\sigma_x,\sigma_y,\sigma_z)$, with the quantization axis along $z$, \begin{equation}
	\mathbf{S}_{\mathbf{k}\alpha} = \frac12 \begin{pmatrix}
		c_{\mathbf{k}\alpha\uparrow}^\dagger & c_{\mathbf{k}\alpha\downarrow}^\dagger
	\end{pmatrix} \boldsymbol{\sigma} \begin{pmatrix}
		c_{\mathbf{k}\alpha\uparrow} \\ c_{\mathbf{k}\alpha\downarrow}
	\end{pmatrix}.
\end{equation} 
We emphasize that the Zeeman term is diagonal in the orbital basis, and we therefore call it an \textit{orbital Zeeman effect} in contrast to the \textit{band Zeeman effect}, which is discussed below. This additional term in the Kane-Mele Hamiltonian $H=H_0^\text{Kane-Mele} + H_\text{HK} + H_\text{orbital Zeeman}$ modifies the ground space. We parameterize the Zeeman field $\mathbf{B}$ in spherical coordinates $(B,\theta,\psi)$ such that \begin{equation}
	\mathbf{B}\cdot\mathbf{S} = B\left[\cos(\theta) S^z + \sin(\theta)\cos(\psi)S^x + \sin(\theta)\sin(\psi) S^y\right].
\end{equation} 
The one-particle Bloch Hamiltonian in the sublattice basis $\lbrace\ket{\mathbf{k}A\uparrow},\ket{\mathbf{k}B\uparrow},\ket{\mathbf{k}A\downarrow},\ket{\mathbf{k}B\downarrow}\rbrace$ is given by \begin{widetext}\begin{equation}\label{eq:hamiltonian_orbital}
	H_0^\text{orbital}(\mathbf{k}) = \begin{pmatrix}
		h_z(\mathbf{k},\uparrow) - B\cos(\theta) & -t\gamma_\mathbf{k}^* & -B\sin(\theta)\text{e}^{-\text{i}\psi} & 0 \\
		-t\gamma_\mathbf{k} & -h_z(\mathbf{k},\uparrow)-B\cos(\theta) & 0 & -B\sin(\theta)\text{e}^{-\text{i}\psi} \\
		-B\sin(\theta)\text{e}^{\text{i}\psi} & 0 & h_z(\mathbf{k},\downarrow) +B\cos(\theta) & -t\gamma_\mathbf{k}^* \\ 0 & -B\sin(\theta)\text{e}^{\text{i}\psi} & -t\gamma_\mathbf{k} & -h_z(\mathbf{k},\downarrow) + B\cos(\theta)
	\end{pmatrix}.
\end{equation}\end{widetext} 
We set $\Delta=0$, so that we have $h_z(\mathbf{k},\uparrow)=-h_z(\mathbf{k},\downarrow)\equiv h_z(\mathbf{k})$. If moreover $\psi=0$, the Hamiltonian can be rewritten in terms of Pauli matrices ($\boldsymbol{\sigma}$ for spin and $\boldsymbol{\tau}$ for the orbital part) like \begin{multline} H_0^\text{orbital}(\mathbf{k}) = h_z(\mathbf{k})\sigma_z\otimes \tau_z -t\left(\mathfrak{Re}(\gamma_\mathbf{k})\tau_x + \mathfrak{Im}(\gamma_\mathbf{k})\tau_y\right) \\ - B\cos(\theta)\sigma_z -B\sin(\theta)\sigma_x.
\end{multline} We define the Bloch chiral symmetry operator $S=\sigma_y\otimes\tau_z$ which anticommutes with the Bloch Hamiltonian.

\begin{figure*}[!ht]
	\centering
	\includegraphics[scale=1]{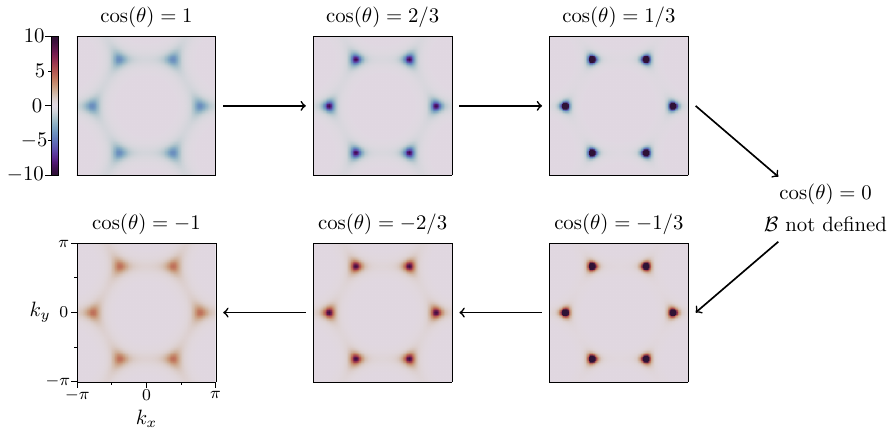}
	\caption{Berry curvatures of the lowest band in the Kane-Mele model with an orbital Zeeman term, with parameters $t=1$, $\Delta=0$, $\varphi=\pi/2$,$t'=0.1$, $B=0.1$ and $\psi=0$. The axes and the colobar are the same for all plot (arbitrary units). The Berry curvature is cut if $\abs{\mathcal{B}}>10$. For $\theta=\pi/2$, the gap closes at the Dirac points.}
	\label{fig:berry_orbital}
\end{figure*}

This model has four distinct bands, and we focus on the lowest band as well as on its Berry curvature. We examine how this Berry curvature evolves as a function of $\theta$, the angle of the Zeeman field relative to the $(xy)$ plane. When $\theta$ is $0$ or $\pi$, the lowest band is only occupied by a single spin species: for $\theta=0$, only the electrons bearing spin $\uparrow$ contribute, and for $\theta= \pi$, only electrons bearing spin $\downarrow$ contribute. For intermediate values of $\theta$, a combination of both spin species contributes. The Berry curvature for the lowest band $\ket{u_0}$ defined as \begin{equation}\label{eq:berry_curvature}
	\mathcal{B}(\mathbf{k}) = \text{i} (\bra{\partial_x u_0} \partial_y u_0 \rangle - \bra{\partial_y u_0} \partial_x u_0 \rangle )
\end{equation} is shown in Fig.~\ref{fig:berry_orbital}, with the parameters set for the topological phase of the Kane-Mele model~\footnote{The computations of Berry curvatures are done for basis II \cite{Bena2009} over the full Brillouin zone. The numerical method used is the one of \cite{Fukui2005}.}. As depicted in this figure, we observe the expected Berry curvature for the two topological Haldane models at $\theta=0$ and $\theta=\pi$ (because we fix the spin to be either $\uparrow$ or $\downarrow$ at these values, these are the original Haldane models giving birth to the Kane-Mele model). Surprinsingly, for intermediate angles, the Berry curvature remains localized at the Dirac points, and its value at these points diverges as $\cos(\theta)$, the projection orthogonal to the $(xy)$ plane, approaches zero.

\begin{figure}[!ht]
	\centering
	\includegraphics[scale=1]{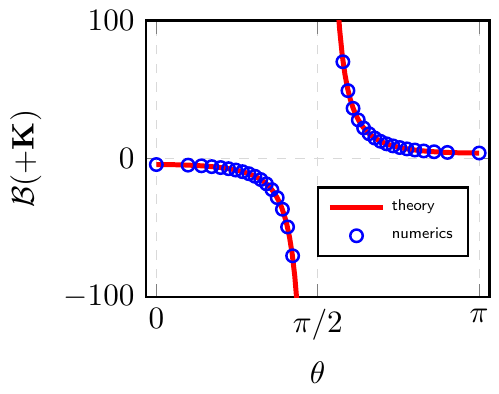}
	\caption{Evolution of the Berry curvature (arbitrary units) at the Dirac point $+\mathbf{K}$ of the lowest band in the Kane-Mele model with an orbital Zeeman term, with parameters $t=1$, $\Delta=0$, $\varphi=\pi/2$,$t'=0.1$, $B=0.1$ and $\psi=0$.}
	\label{fig:berry_curvature_dirac_point}
\end{figure}

We confirm this observation by exploiting the chiral symmetry (for details, see App.~\ref{app:chiral}), i.e., the existence of a Hermitian operator $S$ that anticommutes with the Hamiltonian, $\{S,H_0^{\text{orbital}}(\textbf{k})\}=0$, to diagonalize the Hamiltonian and compute the Berry curvature at the Dirac points $\xi\mathbf{K}$ with the valley index $\xi\in\lbrace -1,1\rbrace$. For the critical value $\mathfrak{h}\equiv -\xi 3\sqrt{3}t'$, which corresponds to the topological transition in Haldane model, we obtain the following expression for the Berry curvature: \begin{equation}
	\mathcal{B}(\xi\mathbf{K}) = -\frac{\text{sgn}\left[\cos(\theta)\right]\hbar^2 v^2}{2\left[\mathfrak{h}\cos(\theta)\right]^2},
\end{equation} with $v=3at/(2\hbar)$, where $a$ is the intersite distance on the honeycomb lattice, and $\text{sgn}=x/|x|$ the sign function. This equation successfully reproduces the behavior observed in numerical simulations, where $\cos(\theta)$ appears in the denominator, as shown in Fig.~\ref{fig:berry_curvature_dirac_point}. 

Now, we turn our attention back to the Hall conductivity. The introduction of the Zeeman field naturally removes the many-body degeneracy, leaving only a single ground state: the Slater determinant of the lowest band. In this system, the Hall conductivity, defined via the velocity operator in Eq.~(\ref{eq:vk}), is computed by integrating the Berry curvature over the first Brillouin zone. For $\theta\neq\pi/2$, the numerical result is, using Ref. \cite{Fukui2005}, \begin{equation}\label{eq:sigma_orbital_zeeman}
	\sigma_{xy}^\text{orbital} = -\text{sgn}\left[\cos(\theta)\right]\frac{e^2}{h}.
\end{equation} 
This band still exhibits a quantized ``Chern number", and more interestingly, the value of the Hall conductivity, also quantized, is determined by the spin species that the Zeeman field favors. When the field points in the upper half-plane, spins $\uparrow$ are preferred, and the Hall conductivity becomes $-1$, matching the value of the independent Haldane model for spins $\uparrow$.

To better understand this point, let us focus on a given $\mathbf{k}$ and examine the situation in the absence of the Zeeman field. Without the field, states $\ket{\mathbf{k}-\!\uparrow}$ and $\ket{\mathbf{k}-\!\downarrow}$ are degenerate in energy. We can treat the Zeeman Hamiltonian as a small perturbation to the Kane-Mele Hamiltonian and apply degenerate perturbation theory. Specifically, we need to diagonalize the restriction of the Zeeman Hamiltonian $H_Z^\text{orbital}$ to the subspace spanned by the two degenerate states $\ket{\mathbf{k}-\!\uparrow}$ and $\ket{\mathbf{k}-\!\downarrow}$. This restriction reads \begin{equation}
	W = B\begin{pmatrix}
		-\cos(\theta) & -\sin(\theta) \abs{t \gamma_\mathbf{k}}/\epsilon_\mathbf{k} \\
		-\sin(\theta) \abs{t \gamma_\mathbf{k}}/\epsilon_\mathbf{k} & \cos(\theta)
	\end{pmatrix},
\end{equation} which vanishes at the Dirac points. This form of the perturbing Zeeman field explicitly shows that the Zeeman term couples the $\uparrow$ and $\downarrow$ Kane-Mele bands (and not just the orbital degrees of freedom).

Assuming that the contribution to the ``Chern number" is generated at the Dirac points, as it is the case for the Kane-Mele model, we examine the behavior near these points. Near the Dirac points, we can approximate $\gamma_\mathbf{k}\approx 0$, leading to the following form for the perturbation \begin{equation}
	\mathbf{k}\approx\pm\mathbf{K},\quad H_0^\text{orbital}(\mathbf{k})\big\vert_\text{bands -} \approx \begin{pmatrix}
		-B\cos(\theta) & 0 \\
		0 & B\cos(\theta)
	\end{pmatrix}.
\end{equation} Thus, depending on the sign of $\cos(\theta)$ (whether $\theta$ is greater or less than $\pi/2$), the Berry curvature from either the original Kane-Mele $\uparrow$ band or of the $\downarrow$ band dominates. This effect is seen in both valleys, because the Hamiltonian does not introduce any valley-dependent terms.

This simple perturbative argument thus replicates the result of Eq.~(\ref{eq:sigma_orbital_zeeman}): the Hall conductivity remains quantized. The result derived by NTW is preserved, even though the typical assumptions do not strictly hold, because the Green's functions of HK models do not vanish at large distances. This orbital Zeeman Hamiltonian does not break the quantization. One could have thought that selecting a mix of the two spin channels could have resulted in a mix of -1 (up) and +1 (down) in the Hall conductivity such that it is not quantized. But instead, the mixing between the spin and the orbital degrees of freedom compensates the selection of the spin channel and the quantization remains.

\subsection{Band Zeeman term}
\label{sec:bandZ}

In this section, we further explore a new ground state by selecting the spin direction at each $\mathbf{k}$, following the intuition about ferromagnetic states which should break the quantization of the Hall conductivity as an unequal mixture of both spin species. Taking the HK interaction and applying a mean-field treatment on it results in a Zeeman term diagonal in band space. This term may in general be $\mathbf{k}$-dependent, as it is given by $U\sigma_z\langle n_{\mathbf{k}-\downarrow} - n_{\mathbf{k}-\uparrow}\rangle/2$ that could have a $\mathbf{k}$-space structure, but for simplicity and illustration purposes, we only consider a $\mathbf{k}$-invariant Zeeman field given by 
\begin{equation}
	H_\text{band Zeeman} = -\mathbf{B}\cdot\sum_{\mathbf{k}\pm} \mathbf{S}_{\mathbf{k}\pm}.
	\end{equation}  
The complete Bloch Hamiltonian in the complete band basis $\lbrace \ket{+\!\uparrow},\ket{+\!\downarrow},\ket{-\!\uparrow},\ket{-\!\downarrow}\rbrace$ is now block-diagonal,
\begin{widetext}\begin{equation}\label{eq:zeeman_band}
	H_0^\text{band}(\mathbf{k}) = \begin{pmatrix}\epsilon_\mathbf{k} - B\cos(\theta) & -B\sin(\theta)\text{e}^{-\text{i}\psi} & 0 & 0 \\
-B\sin(\theta)\text{e}^{\text{i}\psi} & \epsilon_\mathbf{k} + B\cos(\theta) & 0 & 0 \\
0 & 0 & -\epsilon_\mathbf{k} - B\cos(\theta) & -B\sin(\theta)\text{e}^{-\text{i}\psi} \\
0 & 0 & -B\sin(\theta)\text{e}^{\text{i}\psi} & -\epsilon_\mathbf{k} + B\cos(\theta)\end{pmatrix}.
\end{equation}\end{widetext} The lowest energy eigenstate corresponds to $E_0(\mathbf{k}) = -\epsilon_\mathbf{k} -B$ for the eigenstate \begin{equation}
	\ket{u_0(\mathbf{k})} = \cos\left(\frac\theta2\right)\ket{\mathbf{k}-\uparrow} + \text{e}^{\text{i}\psi}\sin\left(\frac\theta2\right)\ket{\mathbf{k}-\downarrow}.
\end{equation} The spin-coordinates no longer depend on $\mathbf{k}$. The ground state is fully spin-polarized along the direction of the Zeeman field $\mathbf{B}$, as intended.

\begin{figure}[!ht]
	\includegraphics[scale=1]{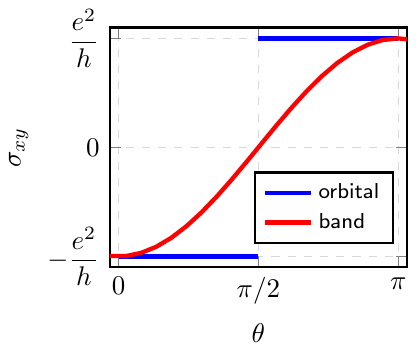}
	\caption{Evolution of the Hall conductivity for the Kane-Mele model with HK interaction at quarter-filling for Zeeman fields diagonal in orbitals or in bands.}
	\label{fig:sigma_xy}
\end{figure}

Analytically, we know that the states $\ket{\mathbf{k}\pm\sigma}$ belong to the two-dimensional vector space $\text{Span}\left(\ket{\mathbf{k}A\sigma},\ket{\mathbf{k}B\sigma}\right)$, meaning that $\ket{\mathbf{k}\pm\uparrow}$ and $\ket{\mathbf{k}'\pm'\downarrow}$ are orthogonal to each other (spin orthogonality). This orthogonality extends to their derivatives as well.  Consequently, we find that \begin{multline}\label{eq:berry_computation}
	\sca{\partial_\mu u_0(\mathbf{k})}{\partial_\nu u_0(\mathbf{k})} = \cos\left(\frac{\theta}2\right)^2\sca{\partial_\mu \mathbf{k}-\uparrow}{\partial_\nu \mathbf{k}-\uparrow} \\ +\sin\left(\frac{\theta}2\right)^2\sca{\partial_\mu \mathbf{k}-\downarrow}{\partial_\nu \mathbf{k}-\downarrow}.
\end{multline} 
Thus, the Berry curvature of the state becomes
 \begin{equation} \label{eq:berry_zeeman_band_analytics}
	\mathcal{B}_0^\text{band}(\mathbf{k}) = \cos\left(\frac{\theta}2\right)^2\mathcal{B}_\uparrow(\mathbf{k}) +\sin\left(\frac{\theta}2\right)^2\mathcal{B}_\downarrow(\mathbf{k}),
\end{equation}
where $\mathcal{B}_\sigma$ are the Berry curvatures inherited from the original Kane-Mele model for the two spin channels.
The Hall conductivity is then the integral of this curvature, which gives, see Fig.~\ref{fig:sigma_xy}, \begin{equation}
	\sigma_{xy}^\text{band} = \left[-\cos\left(\frac\theta2\right)^2 + \sin\left(\frac\theta2\right)^2\right]\frac{e^2}{h}=-\cos(\theta)\frac{e^2}{h}.
\end{equation} This result is no longer quantized (we detail in App.~\ref{app:int_berry} the reasons why the integral of a Berry curvature as defined in Eq.~(\ref{eq:berry_curvature}) can be a non-integer), even though it remains well-defined in this situation. To compare with the result of NTW, we note that it is not the HK interaction that directly leads to this de-quantization, as quantization is preserved for the orbital Zeeman field. Indeed, in the case of the orbital Zeeman term, the eigenstates of the conduction and valence bands, respectively $\lambda=+$ and $-$, get mixed. However, for the band Zeeman field, mixing takes place only between eigenstates of different spin orientation in the same (valence or conduction) band, and the system can still be interpreted as independent uncoupled subsystems. 

Notice that the band Zeeman field may seem somewhat unphysical as it couples states at arbitrary distance, distinguishing it from the previous case where the Zeeman term was local. However, this is inherited from the original HK interaction and its infinite interaction range and the homogeneous order parameter $\langle n_{\mathbf{k}-\downarrow} - n_{\mathbf{k}-\uparrow}\rangle$ that enters the expression of the band Zeeman term if we consider a ferromagnetic ordering. Both the infinite-range interaction and the extensive (albeit not exponential) ground-state degeneracy of the ferromagnetic states $(N+1)$, represented by the different orientations of the field $\mathbf{B}$, break the assumptions made in NTW, ultimately leading to a non-quantized Hall conductivity.

%%%%%%%%%
\section{Conclusion}\label{sec:conc}

In conclusion, we have investigated topological band models, concentrating mainly on the Kane-Mele model, with an infinite-range HK interaction at quarter filling, i.e., when the valence-band spin doublet is half-filled. The original ground-state degeneracy $g_\text{HK}=2^N$, in the case of $N$ spin-$1/2$ electrons, is an artefact of the model -- it is immediately lifted by infinitesimal couplings beyond the pure HK interaction. Here, we consider that the spins get ferromagnetically aligned as a consequence of the reported divergent spin susceptibility \cite{Mai2023a}. The ferromagnetic ordering drastically lowers the original degeneracy to $g_\text{FM}=N+1$, but the latter remains extensive associated with the $z$-component of the magnetization $-N\leq \langle S^z\rangle=N\cos(\theta)\leq N$. If we consider the Hall response, under the above-mentioned omission of contributions from the unphysical infinite-range HK interaction, there is \textit{a priori} no reason to expect a quantized value, as stipulated by the NTW theorem according to which $\sigma_{xy}=n\times e^2/h$ with $n$ an integer value, because of (i) precisely the infinite range of the HK interaction and (ii) the macroscopic ground-state degeneracy $g_\text{FM}$. 

To investigate this Hall conductivity in more detail, we have considered an external Zeeman field that mimics to some extent the ferromagnetic ordering on the mean-field level. Quite generally we have considered two variants of the Zeeman term. The first one is diagonal in the atomic-orbital basis, as one would expect for a natural Zeeman effect, while the second one is diagonal in the band indices, as stipulated by the ferromagnetic order parameter $\langle n_{\mathbf{k}-\downarrow} - n_{\mathbf{k}-\uparrow}\rangle$. Most strikingly, the Hall response is different in the two cases. While in the case of the orbital Zeeman effect, we obtain a quantized Hall response
\begin{equation}
    \sigma_{xy}^\text{orbital}=-\text{sgn}\left[\cos(\theta)\right]\frac{e^2}{h},
\end{equation}
depending only on the sign of the projected spin, the Hall response for a band Zeeman effect that mimics the ferromagnetic order in a direction given by the angle $\theta$ is
\begin{equation}
	\sigma_{xy}^\text{band} =-\cos(\theta)\frac{e^2}{h}=-\frac{\langle S^z\rangle}{N}\frac{e^2}{h},
\end{equation}
 which varies continuously with the projected spin.

This last result can be understood qualitatively if we consider an artificial ``two-step'' diagonalization of the models. The first diagonalization concerns the model in the absence of any Zeeman effect (orbital diagonalization). In this case we obtain two spin doublets, one for the conduction band ($\lambda=+$) and one for the valence band ($\lambda=-$). The corresponding subspaces are spanned by the states $|\textbf{k},\lambda=+,\sigma\rangle$ and $|\textbf{k},\lambda=-,\sigma\rangle$, respectively. The second diagonalization is that of the additional Zeeman term (spin diagonalization). For the situation with a band Zeeman field, it is possible to look directly at the block-diagonal Hamiltonian in Eq.~(\ref{eq:zeeman_band}) and just diagonalize the Zeeman term. Since the two spin branches of each band are associated, in the topological sector, with opposite Chern numbers inherited from the first diagonalization, the Chern number of a state with a magnetization given by the angle $\theta$ is just the weighted sum of both contributions, and the Hall response is thus not quantized. 

The situation is different for the orbital Zeeman effect, which cannot be restricted to the band subspaces so that the two steps of diagonalization do not commute. One therefore needs to diagonalize the full Hamiltonian in Eq.~(\ref{eq:hamiltonian_orbital}), along with the HK interaction, in a single step. In other words, the orbital Zeeman term mixes all four states $\ket{\textbf{k}\lambda\sigma}$ in each $\textbf{k}$ sector. The bands, once isolated have therefore the usual discrete Chern numbers $\mathcal{C}\in \mathbf{Z}$, and the Hall response is quantized.

Notice finally that we do not consider here any magnetic effects induced by the different Zeeman effects, though we would like to mention that the magnetic Kane-Mele model has been recently studied in \cite{Ozaki2023} and \cite{Soni2024}. Beyond our work, significant attention has also been put in the $\mathbf{Z}_2$ invariant \cite{Wang2012a}, which has been simplified and applied to interacting systems \cite{Sinha2025}.

\smallskip

\section*{Acknowledgments}

We would like to thank Olivier Fayet, Axel Guille, Fr\'ed\'eric Pi\'echon, Florian Simon and Alessandro Toschi for fruitful discussions. This work was financially supported by a PhD scholarship from \'Ecole polytechnique (AMX).

\nocite{*}
\bibliography{ref}

\begin{thebibliography}{45}%
\makeatletter
\providecommand \@ifxundefined [1]{%
 \@ifx{#1\undefined}
}%
\providecommand \@ifnum [1]{%
 \ifnum #1\expandafter \@firstoftwo
 \else \expandafter \@secondoftwo
 \fi
}%
\providecommand \@ifx [1]{%
 \ifx #1\expandafter \@firstoftwo
 \else \expandafter \@secondoftwo
 \fi
}%
\providecommand \natexlab [1]{#1}%
\providecommand \enquote  [1]{``#1''}%
\providecommand \bibnamefont  [1]{#1}%
\providecommand \bibfnamefont [1]{#1}%
\providecommand \citenamefont [1]{#1}%
\providecommand \href@noop [0]{\@secondoftwo}%
\providecommand \href [0]{\begingroup \@sanitize@url \@href}%
\providecommand \@href[1]{\@@startlink{#1}\@@href}%
\providecommand \@@href[1]{\endgroup#1\@@endlink}%
\providecommand \@sanitize@url [0]{\catcode `\\12\catcode `\$12\catcode
  `\&12\catcode `\#12\catcode `\^12\catcode `\_12\catcode `\%12\relax}%
\providecommand \@@startlink[1]{}%
\providecommand \@@endlink[0]{}%
\providecommand \url  [0]{\begingroup\@sanitize@url \@url }%
\providecommand \@url [1]{\endgroup\@href {#1}{\urlprefix }}%
\providecommand \urlprefix  [0]{URL }%
\providecommand \Eprint [0]{\href }%
\providecommand \doibase [0]{https://doi.org/}%
\providecommand \selectlanguage [0]{\@gobble}%
\providecommand \bibinfo  [0]{\@secondoftwo}%
\providecommand \bibfield  [0]{\@secondoftwo}%
\providecommand \translation [1]{[#1]}%
\providecommand \BibitemOpen [0]{}%
\providecommand \bibitemStop [0]{}%
\providecommand \bibitemNoStop [0]{.\EOS\space}%
\providecommand \EOS [0]{\spacefactor3000\relax}%
\providecommand \BibitemShut  [1]{\csname bibitem#1\endcsname}%
\let\auto@bib@innerbib\@empty
%</preamble>
\bibitem [{\citenamefont {Cayssol}\ and\ \citenamefont
  {Fuchs}(2021)}]{Cayssol2021}%
  \BibitemOpen
  \bibfield  {author} {\bibinfo {author} {\bibfnamefont {J.}~\bibnamefont
  {Cayssol}}\ and\ \bibinfo {author} {\bibfnamefont {J.~N.}\ \bibnamefont
  {Fuchs}},\ }\bibfield  {title} {\bibinfo {title} {Topological and geometrical
  aspects of band theory},\ }\href {https://doi.org/10.1088/2515-7639/abf0b5}
  {\bibfield  {journal} {\bibinfo  {journal} {J. Phys. Mater.}\ }\textbf
  {\bibinfo {volume} {4}},\ \bibinfo {pages} {034007} (\bibinfo {year}
  {2021})}\BibitemShut {NoStop}%
\bibitem [{\citenamefont {Thouless}\ \emph {et~al.}(1982)\citenamefont
  {Thouless}, \citenamefont {Kohmoto}, \citenamefont {Nightingale},\ and\
  \citenamefont {Den~Nijs}}]{Thouless1982}%
  \BibitemOpen
  \bibfield  {author} {\bibinfo {author} {\bibfnamefont {D.~J.}\ \bibnamefont
  {Thouless}}, \bibinfo {author} {\bibfnamefont {M.}~\bibnamefont {Kohmoto}},
  \bibinfo {author} {\bibfnamefont {M.~P.}\ \bibnamefont {Nightingale}},\ and\
  \bibinfo {author} {\bibfnamefont {M.}~\bibnamefont {Den~Nijs}},\ }\bibfield
  {title} {\bibinfo {title} {Quantized {{Hall Conductance}} in a
  {{Two-Dimensional Periodic Potential}}},\ }\href
  {https://doi.org/10.1103/PhysRevLett.49.405} {\bibfield  {journal} {\bibinfo
  {journal} {Phys. Rev. Lett.}\ }\textbf {\bibinfo {volume} {49}},\ \bibinfo
  {pages} {405} (\bibinfo {year} {1982})}\BibitemShut {NoStop}%
\bibitem [{\citenamefont {Haldane}(1988)}]{Haldane1988}%
  \BibitemOpen
  \bibfield  {author} {\bibinfo {author} {\bibfnamefont {F.~D.~M.}\
  \bibnamefont {Haldane}},\ }\bibfield  {title} {\bibinfo {title} {Model for a
  {{Quantum Hall Effect}} without {{Landau Levels}}: {{Condensed-Matter
  Realization}} of the "{{Parity Anomaly}}"},\ }\href
  {https://doi.org/10.1103/PhysRevLett.61.2015} {\bibfield  {journal} {\bibinfo
   {journal} {Phys. Rev. Lett.}\ }\textbf {\bibinfo {volume} {61}},\ \bibinfo
  {pages} {2015} (\bibinfo {year} {1988})}\BibitemShut {NoStop}%
\bibitem [{\citenamefont {v.~Klitzing}\ \emph {et~al.}(1980)\citenamefont
  {v.~Klitzing}, \citenamefont {Dorda},\ and\ \citenamefont
  {Pepper}}]{Klitzing1980}%
  \BibitemOpen
  \bibfield  {author} {\bibinfo {author} {\bibfnamefont {K.}~\bibnamefont
  {v.~Klitzing}}, \bibinfo {author} {\bibfnamefont {G.}~\bibnamefont {Dorda}},\
  and\ \bibinfo {author} {\bibfnamefont {M.}~\bibnamefont {Pepper}},\
  }\bibfield  {title} {\bibinfo {title} {New {{Method}} for {{High-Accuracy
  Determination}} of the {{Fine-Structure Constant Based}} on {{Quantized Hall
  Resistance}}},\ }\href {https://doi.org/10.1103/PhysRevLett.45.494}
  {\bibfield  {journal} {\bibinfo  {journal} {Phys. Rev. Lett.}\ }\textbf
  {\bibinfo {volume} {45}},\ \bibinfo {pages} {494} (\bibinfo {year}
  {1980})}\BibitemShut {NoStop}%
\bibitem [{\citenamefont {Niu}\ \emph {et~al.}(1985)\citenamefont {Niu},
  \citenamefont {Thouless},\ and\ \citenamefont {Wu}}]{Niu1985}%
  \BibitemOpen
  \bibfield  {author} {\bibinfo {author} {\bibfnamefont {Q.}~\bibnamefont
  {Niu}}, \bibinfo {author} {\bibfnamefont {D.~J.}\ \bibnamefont {Thouless}},\
  and\ \bibinfo {author} {\bibfnamefont {Y.-S.}\ \bibnamefont {Wu}},\
  }\bibfield  {title} {\bibinfo {title} {Quantized {{Hall}} conductance as a
  topological invariant},\ }\href {https://doi.org/10.1103/PhysRevB.31.3372}
  {\bibfield  {journal} {\bibinfo  {journal} {Phys. Rev. B}\ }\textbf {\bibinfo
  {volume} {31}},\ \bibinfo {pages} {3372} (\bibinfo {year}
  {1985})}\BibitemShut {NoStop}%
\bibitem [{\citenamefont {Kudo}\ \emph {et~al.}(2019)\citenamefont {Kudo},
  \citenamefont {Watanabe}, \citenamefont {Kariyado},\ and\ \citenamefont
  {Hatsugai}}]{Kudo2019}%
  \BibitemOpen
  \bibfield  {author} {\bibinfo {author} {\bibfnamefont {K.}~\bibnamefont
  {Kudo}}, \bibinfo {author} {\bibfnamefont {H.}~\bibnamefont {Watanabe}},
  \bibinfo {author} {\bibfnamefont {T.}~\bibnamefont {Kariyado}},\ and\
  \bibinfo {author} {\bibfnamefont {Y.}~\bibnamefont {Hatsugai}},\ }\bibfield
  {title} {\bibinfo {title} {Many-{{Body Chern Number}} without
  {{Integration}}},\ }\href {https://doi.org/10.1103/PhysRevLett.122.146601}
  {\bibfield  {journal} {\bibinfo  {journal} {Phys. Rev. Lett.}\ }\textbf
  {\bibinfo {volume} {122}},\ \bibinfo {pages} {146601} (\bibinfo {year}
  {2019})}\BibitemShut {NoStop}%
\bibitem [{\citenamefont {Goldman}\ and\ \citenamefont
  {Ozawa}(2024)}]{Goldman2024}%
  \BibitemOpen
  \bibfield  {author} {\bibinfo {author} {\bibfnamefont {N.}~\bibnamefont
  {Goldman}}\ and\ \bibinfo {author} {\bibfnamefont {T.}~\bibnamefont
  {Ozawa}},\ }\bibfield  {title} {\bibinfo {title} {Relating the {{Hall}}
  conductivity to the many-body {{Chern}} number using {{Fermi}}'s {{Golden}}
  rule and {{Kramers}}--{{Kronig}} relations},\ }\href
  {https://doi.org/10.5802/crphys.191} {\bibfield  {journal} {\bibinfo
  {journal} {Comptes Rendus. Physique}\ }\textbf {\bibinfo {volume} {25}},\
  \bibinfo {pages} {289} (\bibinfo {year} {2024})}\BibitemShut {NoStop}%
\bibitem [{\citenamefont {Volovik}(2003)}]{Volovik2003}%
  \BibitemOpen
  \bibfield  {author} {\bibinfo {author} {\bibfnamefont {G.~E.}\ \bibnamefont
  {Volovik}},\ }\href@noop {} {\emph {\bibinfo {title} {{{THE UNIVERSE IN A
  HELIUM DROPLET}}}}},\ International {{Series}} of {{Monographs}} on
  {{Physics}}\ (\bibinfo  {publisher} {Oxford University Press},\ \bibinfo
  {address} {Oxford},\ \bibinfo {year} {2003})\BibitemShut {NoStop}%
\bibitem [{\citenamefont {Wang}\ \emph {et~al.}(2010)\citenamefont {Wang},
  \citenamefont {Qi},\ and\ \citenamefont {Zhang}}]{Wang2010}%
  \BibitemOpen
  \bibfield  {author} {\bibinfo {author} {\bibfnamefont {Z.}~\bibnamefont
  {Wang}}, \bibinfo {author} {\bibfnamefont {X.-L.}\ \bibnamefont {Qi}},\ and\
  \bibinfo {author} {\bibfnamefont {S.-C.}\ \bibnamefont {Zhang}},\ }\bibfield
  {title} {\bibinfo {title} {Topological {{Order Parameters}} for {{Interacting
  Topological Insulators}}},\ }\href
  {https://doi.org/10.1103/PhysRevLett.105.256803} {\bibfield  {journal}
  {\bibinfo  {journal} {Phys. Rev. Lett.}\ }\textbf {\bibinfo {volume} {105}},\
  \bibinfo {pages} {256803} (\bibinfo {year} {2010})}\BibitemShut {NoStop}%
\bibitem [{\citenamefont {Wang}\ and\ \citenamefont {Zhang}(2012)}]{Wang2012}%
  \BibitemOpen
  \bibfield  {author} {\bibinfo {author} {\bibfnamefont {Z.}~\bibnamefont
  {Wang}}\ and\ \bibinfo {author} {\bibfnamefont {S.-C.}\ \bibnamefont
  {Zhang}},\ }\bibfield  {title} {\bibinfo {title} {Simplified {{Topological
  Invariants}} for {{Interacting Insulators}}},\ }\href
  {https://doi.org/10.1103/PhysRevX.2.031008} {\bibfield  {journal} {\bibinfo
  {journal} {Phys. Rev. X}\ }\textbf {\bibinfo {volume} {2}},\ \bibinfo {pages}
  {031008} (\bibinfo {year} {2012})}\BibitemShut {NoStop}%
\bibitem [{\citenamefont {Peralta~Gavensky}\ \emph {et~al.}(2023)\citenamefont
  {Peralta~Gavensky}, \citenamefont {Sachdev},\ and\ \citenamefont
  {Goldman}}]{PeraltaGavensky2023}%
  \BibitemOpen
  \bibfield  {author} {\bibinfo {author} {\bibfnamefont {L.}~\bibnamefont
  {Peralta~Gavensky}}, \bibinfo {author} {\bibfnamefont {S.}~\bibnamefont
  {Sachdev}},\ and\ \bibinfo {author} {\bibfnamefont {N.}~\bibnamefont
  {Goldman}},\ }\bibfield  {title} {\bibinfo {title} {Connecting the
  {{Many-Body Chern Number}} to {{Luttinger}}'s {{Theorem}} through {{St\v
  reda}}'s {{Formula}}},\ }\href
  {https://doi.org/10.1103/PhysRevLett.131.236601} {\bibfield  {journal}
  {\bibinfo  {journal} {Phys. Rev. Lett.}\ }\textbf {\bibinfo {volume} {131}},\
  \bibinfo {pages} {236601} (\bibinfo {year} {2023})}\BibitemShut {NoStop}%
\bibitem [{\citenamefont {Zhao}\ \emph {et~al.}(2023)\citenamefont {Zhao},
  \citenamefont {Mai}, \citenamefont {Bradlyn},\ and\ \citenamefont
  {Phillips}}]{Zhao2023}%
  \BibitemOpen
  \bibfield  {author} {\bibinfo {author} {\bibfnamefont {J.}~\bibnamefont
  {Zhao}}, \bibinfo {author} {\bibfnamefont {P.}~\bibnamefont {Mai}}, \bibinfo
  {author} {\bibfnamefont {B.}~\bibnamefont {Bradlyn}},\ and\ \bibinfo {author}
  {\bibfnamefont {P.}~\bibnamefont {Phillips}},\ }\bibfield  {title} {\bibinfo
  {title} {Failure of {{Topological Invariants}} in {{Strongly Correlated
  Matter}}},\ }\href {https://doi.org/10.1103/PhysRevLett.131.106601}
  {\bibfield  {journal} {\bibinfo  {journal} {Phys. Rev. Lett.}\ }\textbf
  {\bibinfo {volume} {131}},\ \bibinfo {pages} {106601} (\bibinfo {year}
  {2023})}\BibitemShut {NoStop}%
\bibitem [{\citenamefont {Yang}(2019)}]{Yang2019}%
  \BibitemOpen
  \bibfield  {author} {\bibinfo {author} {\bibfnamefont {M.-F.}\ \bibnamefont
  {Yang}},\ }\bibfield  {title} {\bibinfo {title} {Manifestation of topological
  behaviors in interacting {{Weyl}} systems: {{One-body}} versus two-body
  correlations},\ }\href {https://doi.org/10.1103/PhysRevB.100.245137}
  {\bibfield  {journal} {\bibinfo  {journal} {Phys. Rev. B}\ }\textbf {\bibinfo
  {volume} {100}},\ \bibinfo {pages} {245137} (\bibinfo {year}
  {2019})}\BibitemShut {NoStop}%
\bibitem [{\citenamefont {Bollmann}\ \emph {et~al.}(2024)\citenamefont
  {Bollmann}, \citenamefont {Setty}, \citenamefont {Seifert},\ and\
  \citenamefont {K{\"o}nig}}]{Bollmann2024}%
  \BibitemOpen
  \bibfield  {author} {\bibinfo {author} {\bibfnamefont {S.}~\bibnamefont
  {Bollmann}}, \bibinfo {author} {\bibfnamefont {C.}~\bibnamefont {Setty}},
  \bibinfo {author} {\bibfnamefont {U.~F.~P.}\ \bibnamefont {Seifert}},\ and\
  \bibinfo {author} {\bibfnamefont {E.~J.}\ \bibnamefont {K{\"o}nig}},\
  }\bibfield  {title} {\bibinfo {title} {Topological {{Green}}'s {{Function
  Zeros}} in an {{Exactly Solved Model}} and {{Beyond}}},\ }\href
  {https://doi.org/10.1103/PhysRevLett.133.136504} {\bibfield  {journal}
  {\bibinfo  {journal} {Phys. Rev. Lett.}\ }\textbf {\bibinfo {volume} {133}},\
  \bibinfo {pages} {136504} (\bibinfo {year} {2024})}\BibitemShut {NoStop}%
\bibitem [{\citenamefont {Gurarie}(2011)}]{Gurarie2011}%
  \BibitemOpen
  \bibfield  {author} {\bibinfo {author} {\bibfnamefont {V.}~\bibnamefont
  {Gurarie}},\ }\bibfield  {title} {\bibinfo {title} {Single-particle
  {{Green}}'s functions and interacting topological insulators},\ }\href
  {https://doi.org/10.1103/PhysRevB.83.085426} {\bibfield  {journal} {\bibinfo
  {journal} {Phys. Rev. B}\ }\textbf {\bibinfo {volume} {83}},\ \bibinfo
  {pages} {085426} (\bibinfo {year} {2011})}\BibitemShut {NoStop}%
\bibitem [{\citenamefont {Wagner}\ \emph {et~al.}(2023)\citenamefont {Wagner},
  \citenamefont {Crippa}, \citenamefont {Amaricci}, \citenamefont {Hansmann},
  \citenamefont {Klett}, \citenamefont {K{\"o}nig}, \citenamefont
  {Sch{\"a}fer}, \citenamefont {Sante}, \citenamefont {Cano}, \citenamefont
  {Millis}, \citenamefont {Georges},\ and\ \citenamefont
  {Sangiovanni}}]{Wagner2023}%
  \BibitemOpen
  \bibfield  {author} {\bibinfo {author} {\bibfnamefont {N.}~\bibnamefont
  {Wagner}}, \bibinfo {author} {\bibfnamefont {L.}~\bibnamefont {Crippa}},
  \bibinfo {author} {\bibfnamefont {A.}~\bibnamefont {Amaricci}}, \bibinfo
  {author} {\bibfnamefont {P.}~\bibnamefont {Hansmann}}, \bibinfo {author}
  {\bibfnamefont {M.}~\bibnamefont {Klett}}, \bibinfo {author} {\bibfnamefont
  {E.~J.}\ \bibnamefont {K{\"o}nig}}, \bibinfo {author} {\bibfnamefont
  {T.}~\bibnamefont {Sch{\"a}fer}}, \bibinfo {author} {\bibfnamefont {D.~D.}\
  \bibnamefont {Sante}}, \bibinfo {author} {\bibfnamefont {J.}~\bibnamefont
  {Cano}}, \bibinfo {author} {\bibfnamefont {A.~J.}\ \bibnamefont {Millis}},
  \bibinfo {author} {\bibfnamefont {A.}~\bibnamefont {Georges}},\ and\ \bibinfo
  {author} {\bibfnamefont {G.}~\bibnamefont {Sangiovanni}},\ }\bibfield
  {title} {\bibinfo {title} {Mott insulators with boundary zeros},\ }\href
  {https://doi.org/10.1038/s41467-023-42773-7} {\bibfield  {journal} {\bibinfo
  {journal} {Nat Commun}\ }\textbf {\bibinfo {volume} {14}},\ \bibinfo {pages}
  {7531} (\bibinfo {year} {2023})}\BibitemShut {NoStop}%
\bibitem [{\citenamefont {Blason}\ and\ \citenamefont
  {Fabrizio}(2023)}]{Blason2023}%
  \BibitemOpen
  \bibfield  {author} {\bibinfo {author} {\bibfnamefont {A.}~\bibnamefont
  {Blason}}\ and\ \bibinfo {author} {\bibfnamefont {M.}~\bibnamefont
  {Fabrizio}},\ }\bibfield  {title} {\bibinfo {title} {Unified role of
  {{Green}}'s function poles and zeros in correlated topological insulators},\
  }\href {https://doi.org/10.1103/PhysRevB.108.125115} {\bibfield  {journal}
  {\bibinfo  {journal} {Phys. Rev. B}\ }\textbf {\bibinfo {volume} {108}},\
  \bibinfo {pages} {125115} (\bibinfo {year} {2023})}\BibitemShut {NoStop}%
\bibitem [{\citenamefont {Setty}\ \emph {et~al.}(2024)\citenamefont {Setty},
  \citenamefont {Xie}, \citenamefont {Sur}, \citenamefont {Chen}, \citenamefont
  {Vergniory},\ and\ \citenamefont {Si}}]{Setty2024}%
  \BibitemOpen
  \bibfield  {author} {\bibinfo {author} {\bibfnamefont {C.}~\bibnamefont
  {Setty}}, \bibinfo {author} {\bibfnamefont {F.}~\bibnamefont {Xie}}, \bibinfo
  {author} {\bibfnamefont {S.}~\bibnamefont {Sur}}, \bibinfo {author}
  {\bibfnamefont {L.}~\bibnamefont {Chen}}, \bibinfo {author} {\bibfnamefont
  {M.~G.}\ \bibnamefont {Vergniory}},\ and\ \bibinfo {author} {\bibfnamefont
  {Q.}~\bibnamefont {Si}},\ }\bibfield  {title} {\bibinfo {title} {Electronic
  properties, correlated topology, and {{Green}}'s function zeros},\ }\href
  {https://doi.org/10.1103/PhysRevResearch.6.033235} {\bibfield  {journal}
  {\bibinfo  {journal} {Phys. Rev. Res.}\ }\textbf {\bibinfo {volume} {6}},\
  \bibinfo {pages} {033235} (\bibinfo {year} {2024})}\BibitemShut {NoStop}%
\bibitem [{\citenamefont {Simon}\ and\ \citenamefont
  {Morice}(2025)}]{Simon2025}%
  \BibitemOpen
  \bibfield  {author} {\bibinfo {author} {\bibfnamefont {F.}~\bibnamefont
  {Simon}}\ and\ \bibinfo {author} {\bibfnamefont {C.}~\bibnamefont {Morice}},\
  }\href {https://doi.org/10.48550/arXiv.2504.00800} {\bibinfo {title}
  {{$\mathbb{Z}_2$} topological invariants from the {{Green}}'s function
  diagonal zeros}} (\bibinfo {year} {2025}),\ \Eprint
  {https://arxiv.org/abs/2504.00800} {arXiv:2504.00800 [cond-mat]} \BibitemShut
  {NoStop}%
\bibitem [{\citenamefont {Wang}\ and\ \citenamefont {Yan}(2013)}]{Wang2013}%
  \BibitemOpen
  \bibfield  {author} {\bibinfo {author} {\bibfnamefont {Z.}~\bibnamefont
  {Wang}}\ and\ \bibinfo {author} {\bibfnamefont {B.}~\bibnamefont {Yan}},\
  }\bibfield  {title} {\bibinfo {title} {Topological {{Hamiltonian}} as an
  exact tool for topological invariants},\ }\href
  {https://doi.org/10.1088/0953-8984/25/15/155601} {\bibfield  {journal}
  {\bibinfo  {journal} {J. Phys.: Condens. Matter}\ }\textbf {\bibinfo {volume}
  {25}},\ \bibinfo {pages} {155601} (\bibinfo {year} {2013})}\BibitemShut
  {NoStop}%
\bibitem [{\citenamefont {Hohenadler}\ and\ \citenamefont
  {Assaad}(2013)}]{Hohenadler2013}%
  \BibitemOpen
  \bibfield  {author} {\bibinfo {author} {\bibfnamefont {M.}~\bibnamefont
  {Hohenadler}}\ and\ \bibinfo {author} {\bibfnamefont {F.~F.}\ \bibnamefont
  {Assaad}},\ }\bibfield  {title} {\bibinfo {title} {Correlation effects in
  two-dimensional topological insulators},\ }\href
  {https://doi.org/10.1088/0953-8984/25/14/143201} {\bibfield  {journal}
  {\bibinfo  {journal} {J. Phys.: Condens. Matter}\ }\textbf {\bibinfo {volume}
  {25}},\ \bibinfo {pages} {143201} (\bibinfo {year} {2013})}\BibitemShut
  {NoStop}%
\bibitem [{\citenamefont {Rachel}(2018)}]{Rachel2018}%
  \BibitemOpen
  \bibfield  {author} {\bibinfo {author} {\bibfnamefont {S.}~\bibnamefont
  {Rachel}},\ }\bibfield  {title} {\bibinfo {title} {Interacting topological
  insulators: A review},\ }\href {https://doi.org/10.1088/1361-6633/aad6a6}
  {\bibfield  {journal} {\bibinfo  {journal} {Rep. Prog. Phys.}\ }\textbf
  {\bibinfo {volume} {81}},\ \bibinfo {pages} {116501} (\bibinfo {year}
  {2018})}\BibitemShut {NoStop}%
\bibitem [{\citenamefont {Hatsugai}\ and\ \citenamefont
  {Kohmoto}(1992)}]{Hatsugai1992}%
  \BibitemOpen
  \bibfield  {author} {\bibinfo {author} {\bibfnamefont {Y.}~\bibnamefont
  {Hatsugai}}\ and\ \bibinfo {author} {\bibfnamefont {M.}~\bibnamefont
  {Kohmoto}},\ }\bibfield  {title} {\bibinfo {title} {Exactly {{Solvable
  Model}} of {{Correlated Lattice Electrons}} in {{Any Dimensions}}},\ }\href
  {https://doi.org/10.1143/JPSJ.61.2056} {\bibfield  {journal} {\bibinfo
  {journal} {J. Phys. Soc. Jpn.}\ }\textbf {\bibinfo {volume} {61}},\ \bibinfo
  {pages} {2056} (\bibinfo {year} {1992})}\BibitemShut {NoStop}%
\bibitem [{\citenamefont {Zhao}\ \emph {et~al.}(2025)\citenamefont {Zhao},
  \citenamefont {Yang},\ and\ \citenamefont {Zhong}}]{Zhao2025}%
  \BibitemOpen
  \bibfield  {author} {\bibinfo {author} {\bibfnamefont {M.}~\bibnamefont
  {Zhao}}, \bibinfo {author} {\bibfnamefont {W.-W.}\ \bibnamefont {Yang}},\
  and\ \bibinfo {author} {\bibfnamefont {Y.}~\bibnamefont {Zhong}},\ }\bibfield
   {title} {\bibinfo {title} {Hatsugai--{{Kohmoto}} models: Exactly solvable
  playground for {{Mottness}} and non-{{Fermi}} liquid},\ }\href
  {https://doi.org/10.1088/1361-648X/adc64c} {\bibfield  {journal} {\bibinfo
  {journal} {J. Phys.: Condens. Matter}\ }\textbf {\bibinfo {volume} {37}},\
  \bibinfo {pages} {183005} (\bibinfo {year} {2025})}\BibitemShut {NoStop}%
\bibitem [{\citenamefont {{Manning-Coe}}\ and\ \citenamefont
  {Bradlyn}(2023)}]{Manning-Coe2023}%
  \BibitemOpen
  \bibfield  {author} {\bibinfo {author} {\bibfnamefont {D.}~\bibnamefont
  {{Manning-Coe}}}\ and\ \bibinfo {author} {\bibfnamefont {B.}~\bibnamefont
  {Bradlyn}},\ }\bibfield  {title} {\bibinfo {title} {Ground state stability,
  symmetry, and degeneracy in {{Mott}} insulators with long-range
  interactions},\ }\href {https://doi.org/10.1103/PhysRevB.108.165136}
  {\bibfield  {journal} {\bibinfo  {journal} {Phys. Rev. B}\ }\textbf {\bibinfo
  {volume} {108}},\ \bibinfo {pages} {165136} (\bibinfo {year}
  {2023})}\BibitemShut {NoStop}%
\bibitem [{\citenamefont {Worm}\ \emph {et~al.}(2024)\citenamefont {Worm},
  \citenamefont {Reitner}, \citenamefont {Held},\ and\ \citenamefont
  {Toschi}}]{Worm2024}%
  \BibitemOpen
  \bibfield  {author} {\bibinfo {author} {\bibfnamefont {P.}~\bibnamefont
  {Worm}}, \bibinfo {author} {\bibfnamefont {M.}~\bibnamefont {Reitner}},
  \bibinfo {author} {\bibfnamefont {K.}~\bibnamefont {Held}},\ and\ \bibinfo
  {author} {\bibfnamefont {A.}~\bibnamefont {Toschi}},\ }\bibfield  {title}
  {\bibinfo {title} {Fermi and {{Luttinger Arcs}}: {{Two Concepts}},
  {{Realized}} on {{One Surface}}},\ }\href
  {https://doi.org/10.1103/PhysRevLett.133.166501} {\bibfield  {journal}
  {\bibinfo  {journal} {Phys. Rev. Lett.}\ }\textbf {\bibinfo {volume} {133}},\
  \bibinfo {pages} {166501} (\bibinfo {year} {2024})}\BibitemShut {NoStop}%
\bibitem [{\citenamefont {Mai}\ \emph {et~al.}(2023{\natexlab{a}})\citenamefont
  {Mai}, \citenamefont {Feldman},\ and\ \citenamefont {Phillips}}]{Mai2023a}%
  \BibitemOpen
  \bibfield  {author} {\bibinfo {author} {\bibfnamefont {P.}~\bibnamefont
  {Mai}}, \bibinfo {author} {\bibfnamefont {B.~E.}\ \bibnamefont {Feldman}},\
  and\ \bibinfo {author} {\bibfnamefont {P.~W.}\ \bibnamefont {Phillips}},\
  }\bibfield  {title} {\bibinfo {title} {Topological {{Mott}} insulator at
  quarter filling in the interacting {{Haldane}} model},\ }\href
  {https://doi.org/10.1103/PhysRevResearch.5.013162} {\bibfield  {journal}
  {\bibinfo  {journal} {Phys. Rev. Res.}\ }\textbf {\bibinfo {volume} {5}},\
  \bibinfo {pages} {013162} (\bibinfo {year} {2023}{\natexlab{a}})}\BibitemShut
  {NoStop}%
\bibitem [{\citenamefont {Mai}\ \emph {et~al.}(2023{\natexlab{b}})\citenamefont
  {Mai}, \citenamefont {Feldman},\ and\ \citenamefont {Phillips}}]{Mai2023}%
  \BibitemOpen
  \bibfield  {author} {\bibinfo {author} {\bibfnamefont {P.}~\bibnamefont
  {Mai}}, \bibinfo {author} {\bibfnamefont {B.~E.}\ \bibnamefont {Feldman}},\
  and\ \bibinfo {author} {\bibfnamefont {P.~W.}\ \bibnamefont {Phillips}},\
  }\bibfield  {title} {\bibinfo {title} {1/4 is the new 1/2 when topology is
  intertwined with {{Mottness}}},\ }\href
  {https://doi.org/10.1038/s41467-023-41465-6} {\bibfield  {journal} {\bibinfo
  {journal} {Nat Commun}\ }\textbf {\bibinfo {volume} {14}},\ \bibinfo {pages}
  {5999} (\bibinfo {year} {2023}{\natexlab{b}})}\BibitemShut {NoStop}%
\bibitem [{\citenamefont {Jab{\l}onowski}\ \emph {et~al.}(2023)\citenamefont
  {Jab{\l}onowski}, \citenamefont {Skolimowski}, \citenamefont {Brzezicki},
  \citenamefont {Byczuk},\ and\ \citenamefont
  {Wysoki{\'n}ski}}]{Jablonowski2023}%
  \BibitemOpen
  \bibfield  {author} {\bibinfo {author} {\bibfnamefont {K.}~\bibnamefont
  {Jab{\l}onowski}}, \bibinfo {author} {\bibfnamefont {J.}~\bibnamefont
  {Skolimowski}}, \bibinfo {author} {\bibfnamefont {W.}~\bibnamefont
  {Brzezicki}}, \bibinfo {author} {\bibfnamefont {K.}~\bibnamefont {Byczuk}},\
  and\ \bibinfo {author} {\bibfnamefont {M.~M.}\ \bibnamefont
  {Wysoki{\'n}ski}},\ }\bibfield  {title} {\bibinfo {title} {Topological
  {{Mott}} insulator in the odd-integer filled {{Anderson}} lattice model with
  {{Hatsugai-Kohmoto}} interactions},\ }\href
  {https://doi.org/10.1103/PhysRevB.108.195145} {\bibfield  {journal} {\bibinfo
   {journal} {Phys. Rev. B}\ }\textbf {\bibinfo {volume} {108}},\ \bibinfo
  {pages} {195145} (\bibinfo {year} {2023})}\BibitemShut {NoStop}%
\bibitem [{\citenamefont {Wysoki{\'n}ski}\ and\ \citenamefont
  {Brzezicki}(2023)}]{Wysokinski2023}%
  \BibitemOpen
  \bibfield  {author} {\bibinfo {author} {\bibfnamefont {M.~M.}\ \bibnamefont
  {Wysoki{\'n}ski}}\ and\ \bibinfo {author} {\bibfnamefont {W.}~\bibnamefont
  {Brzezicki}},\ }\bibfield  {title} {\bibinfo {title} {Quantum anomalous
  {{Hall}} insulator in ionic {{Rashba}} lattice of correlated electrons},\
  }\href {https://doi.org/10.1103/PhysRevB.108.035121} {\bibfield  {journal}
  {\bibinfo  {journal} {Phys. Rev. B}\ }\textbf {\bibinfo {volume} {108}},\
  \bibinfo {pages} {035121} (\bibinfo {year} {2023})}\BibitemShut {NoStop}%
\bibitem [{\citenamefont {Mai}\ \emph {et~al.}(2024)\citenamefont {Mai},
  \citenamefont {Zhao}, \citenamefont {Tenkila}, \citenamefont {Hackner},
  \citenamefont {Kush}, \citenamefont {Pan},\ and\ \citenamefont
  {Phillips}}]{Mai2024}%
  \BibitemOpen
  \bibfield  {author} {\bibinfo {author} {\bibfnamefont {P.}~\bibnamefont
  {Mai}}, \bibinfo {author} {\bibfnamefont {J.}~\bibnamefont {Zhao}}, \bibinfo
  {author} {\bibfnamefont {G.}~\bibnamefont {Tenkila}}, \bibinfo {author}
  {\bibfnamefont {N.~A.}\ \bibnamefont {Hackner}}, \bibinfo {author}
  {\bibfnamefont {D.}~\bibnamefont {Kush}}, \bibinfo {author} {\bibfnamefont
  {D.}~\bibnamefont {Pan}},\ and\ \bibinfo {author} {\bibfnamefont {P.~W.}\
  \bibnamefont {Phillips}},\ }\href {https://doi.org/10.48550/arXiv.2401.08746}
  {\bibinfo {title} {New {{Approach}} to {{Strong Correlation}}: {{Twisting
  Hubbard}} into the {{Orbital Hatsugai-Kohmoto Model}}}} (\bibinfo {year}
  {2024}),\ \Eprint {https://arxiv.org/abs/2401.08746} {arXiv:2401.08746
  [cond-mat]} \BibitemShut {NoStop}%
\bibitem [{\citenamefont {Skolimowski}\ and\ \citenamefont
  {Brzezicki}(2025)}]{Skolimowski2025}%
  \BibitemOpen
  \bibfield  {author} {\bibinfo {author} {\bibfnamefont {J.}~\bibnamefont
  {Skolimowski}}\ and\ \bibinfo {author} {\bibfnamefont {W.}~\bibnamefont
  {Brzezicki}},\ }\bibfield  {title} {\bibinfo {title} {Fate of gapless edge
  states in two-dimensional topological insulators with {{Hatsugai-Kohmoto}}
  interaction},\ }\href {https://doi.org/10.1103/PhysRevB.111.125135}
  {\bibfield  {journal} {\bibinfo  {journal} {Phys. Rev. B}\ }\textbf {\bibinfo
  {volume} {111}},\ \bibinfo {pages} {125135} (\bibinfo {year}
  {2025})}\BibitemShut {NoStop}%
\bibitem [{\citenamefont {Kane}\ and\ \citenamefont {Mele}(2005)}]{Kane2005}%
  \BibitemOpen
  \bibfield  {author} {\bibinfo {author} {\bibfnamefont {C.~L.}\ \bibnamefont
  {Kane}}\ and\ \bibinfo {author} {\bibfnamefont {E.~J.}\ \bibnamefont
  {Mele}},\ }\bibfield  {title} {\bibinfo {title} {Quantum {{Spin Hall Effect}}
  in {{Graphene}}},\ }\href {https://doi.org/10.1103/PhysRevLett.95.226801}
  {\bibfield  {journal} {\bibinfo  {journal} {Phys. Rev. Lett.}\ }\textbf
  {\bibinfo {volume} {95}},\ \bibinfo {pages} {226801} (\bibinfo {year}
  {2005})}\BibitemShut {NoStop}%
\bibitem [{\citenamefont {Skolimowski}(2024)}]{Skolimowski2024}%
  \BibitemOpen
  \bibfield  {author} {\bibinfo {author} {\bibfnamefont {J.}~\bibnamefont
  {Skolimowski}},\ }\bibfield  {title} {\bibinfo {title} {Real-space analysis
  of {{Hatsugai-Kohmoto}} interaction},\ }\href
  {https://doi.org/10.1103/PhysRevB.109.165129} {\bibfield  {journal} {\bibinfo
   {journal} {Phys. Rev. B}\ }\textbf {\bibinfo {volume} {109}},\ \bibinfo
  {pages} {165129} (\bibinfo {year} {2024})}\BibitemShut {NoStop}%
\bibitem [{\citenamefont {Guerci}\ \emph {et~al.}(2025)\citenamefont {Guerci},
  \citenamefont {Sangiovanni}, \citenamefont {Millis},\ and\ \citenamefont
  {Fabrizio}}]{Guerci2025}%
  \BibitemOpen
  \bibfield  {author} {\bibinfo {author} {\bibfnamefont {D.}~\bibnamefont
  {Guerci}}, \bibinfo {author} {\bibfnamefont {G.}~\bibnamefont {Sangiovanni}},
  \bibinfo {author} {\bibfnamefont {A.~J.}\ \bibnamefont {Millis}},\ and\
  \bibinfo {author} {\bibfnamefont {M.}~\bibnamefont {Fabrizio}},\ }\bibfield
  {title} {\bibinfo {title} {Electrical transport in the {{Hatsugai-Kohmoto}}
  model},\ }\href {https://doi.org/10.1103/PhysRevB.111.075124} {\bibfield
  {journal} {\bibinfo  {journal} {Phys. Rev. B}\ }\textbf {\bibinfo {volume}
  {111}},\ \bibinfo {pages} {075124} (\bibinfo {year} {2025})}\BibitemShut
  {NoStop}%
\bibitem [{\citenamefont {Ma}\ \emph {et~al.}(2025)\citenamefont {Ma},
  \citenamefont {Zhao}, \citenamefont {Huang}, \citenamefont {Kush},
  \citenamefont {Bradlyn},\ and\ \citenamefont {Phillips}}]{Ma2025}%
  \BibitemOpen
  \bibfield  {author} {\bibinfo {author} {\bibfnamefont {Y.}~\bibnamefont
  {Ma}}, \bibinfo {author} {\bibfnamefont {J.}~\bibnamefont {Zhao}}, \bibinfo
  {author} {\bibfnamefont {E.~W.}\ \bibnamefont {Huang}}, \bibinfo {author}
  {\bibfnamefont {D.}~\bibnamefont {Kush}}, \bibinfo {author} {\bibfnamefont
  {B.}~\bibnamefont {Bradlyn}},\ and\ \bibinfo {author} {\bibfnamefont {P.~W.}\
  \bibnamefont {Phillips}},\ }\bibfield  {title} {\bibinfo {title} {Charge
  susceptibility and {{Kubo}} response in {{Hatsugai-Kohmoto-related}}
  models},\ }\href {https://doi.org/10.1103/r4q9-hm45} {\bibfield  {journal}
  {\bibinfo  {journal} {Phys. Rev. B}\ }\textbf {\bibinfo {volume} {112}},\
  \bibinfo {pages} {045109} (\bibinfo {year} {2025})}\BibitemShut {NoStop}%
\bibitem [{Note1()}]{Note1}%
  \BibitemOpen
  \bibinfo {note} {We are indepted to Alessandro Toschi who coined this term
  during common discussions.}\BibitemShut {Stop}%
\bibitem [{\citenamefont {Tzeng}\ \emph {et~al.}(2023)\citenamefont {Tzeng},
  \citenamefont {Chang},\ and\ \citenamefont {Yang}}]{Tzeng2023}%
  \BibitemOpen
  \bibfield  {author} {\bibinfo {author} {\bibfnamefont {Y.-C.}\ \bibnamefont
  {Tzeng}}, \bibinfo {author} {\bibfnamefont {P.-Y.}\ \bibnamefont {Chang}},\
  and\ \bibinfo {author} {\bibfnamefont {M.-F.}\ \bibnamefont {Yang}},\
  }\bibfield  {title} {\bibinfo {title} {Interaction-induced metal to
  topological insulator transition},\ }\href
  {https://doi.org/10.1103/PhysRevB.107.155106} {\bibfield  {journal} {\bibinfo
   {journal} {Phys. Rev. B}\ }\textbf {\bibinfo {volume} {107}},\ \bibinfo
  {pages} {155106} (\bibinfo {year} {2023})}\BibitemShut {NoStop}%
\bibitem [{Note2()}]{Note2}%
  \BibitemOpen
  \bibinfo {note} {The computations of Berry curvatures are done for basis II
  \cite {Bena2009} over the full Brillouin zone. The numerical method used is
  the one of \cite {Fukui2005}.}\BibitemShut {Stop}%
\bibitem [{\citenamefont {Fukui}\ \emph {et~al.}(2005)\citenamefont {Fukui},
  \citenamefont {Hatsugai},\ and\ \citenamefont {Suzuki}}]{Fukui2005}%
  \BibitemOpen
  \bibfield  {author} {\bibinfo {author} {\bibfnamefont {T.}~\bibnamefont
  {Fukui}}, \bibinfo {author} {\bibfnamefont {Y.}~\bibnamefont {Hatsugai}},\
  and\ \bibinfo {author} {\bibfnamefont {H.}~\bibnamefont {Suzuki}},\
  }\bibfield  {title} {\bibinfo {title} {Chern {{Numbers}} in {{Discretized
  Brillouin Zone}}: {{Efficient Method}} of {{Computing}} ({{Spin}}) {{Hall
  Conductances}}},\ }\href {https://doi.org/10.1143/JPSJ.74.1674} {\bibfield
  {journal} {\bibinfo  {journal} {J. Phys. Soc. Jpn.}\ }\textbf {\bibinfo
  {volume} {74}},\ \bibinfo {pages} {1674} (\bibinfo {year}
  {2005})}\BibitemShut {NoStop}%
\bibitem [{\citenamefont {Ozaki}\ and\ \citenamefont
  {Ogata}(2023)}]{Ozaki2023}%
  \BibitemOpen
  \bibfield  {author} {\bibinfo {author} {\bibfnamefont {S.}~\bibnamefont
  {Ozaki}}\ and\ \bibinfo {author} {\bibfnamefont {M.}~\bibnamefont {Ogata}},\
  }\bibfield  {title} {\bibinfo {title} {Topological contribution to magnetism
  in the {{Kane-Mele}} model: {{An}} explicit wave-function approach},\ }\href
  {https://doi.org/10.1103/PhysRevB.107.085201} {\bibfield  {journal} {\bibinfo
   {journal} {Phys. Rev. B}\ }\textbf {\bibinfo {volume} {107}},\ \bibinfo
  {pages} {085201} (\bibinfo {year} {2023})}\BibitemShut {NoStop}%
\bibitem [{\citenamefont {Soni}\ \emph {et~al.}(2024)\citenamefont {Soni},
  \citenamefont {Radhakrishnan}, \citenamefont {Rosenow}, \citenamefont
  {Alvarez},\ and\ \citenamefont {Del~Maestro}}]{Soni2024}%
  \BibitemOpen
  \bibfield  {author} {\bibinfo {author} {\bibfnamefont {R.}~\bibnamefont
  {Soni}}, \bibinfo {author} {\bibfnamefont {H.}~\bibnamefont {Radhakrishnan}},
  \bibinfo {author} {\bibfnamefont {B.}~\bibnamefont {Rosenow}}, \bibinfo
  {author} {\bibfnamefont {G.}~\bibnamefont {Alvarez}},\ and\ \bibinfo {author}
  {\bibfnamefont {A.}~\bibnamefont {Del~Maestro}},\ }\bibfield  {title}
  {\bibinfo {title} {Topological and magnetic properties of the interacting
  {{Bernevig-Hughes-Zhang}} model},\ }\href
  {https://doi.org/10.1103/PhysRevB.109.245115} {\bibfield  {journal} {\bibinfo
   {journal} {Phys. Rev. B}\ }\textbf {\bibinfo {volume} {109}},\ \bibinfo
  {pages} {245115} (\bibinfo {year} {2024})}\BibitemShut {NoStop}%
\bibitem [{\citenamefont {Wang}\ \emph {et~al.}(2012)\citenamefont {Wang},
  \citenamefont {Qi},\ and\ \citenamefont {Zhang}}]{Wang2012a}%
  \BibitemOpen
  \bibfield  {author} {\bibinfo {author} {\bibfnamefont {Z.}~\bibnamefont
  {Wang}}, \bibinfo {author} {\bibfnamefont {X.-L.}\ \bibnamefont {Qi}},\ and\
  \bibinfo {author} {\bibfnamefont {S.-C.}\ \bibnamefont {Zhang}},\ }\bibfield
  {title} {\bibinfo {title} {Topological invariants for interacting topological
  insulators with inversion symmetry},\ }\href
  {https://doi.org/10.1103/PhysRevB.85.165126} {\bibfield  {journal} {\bibinfo
  {journal} {Phys. Rev. B}\ }\textbf {\bibinfo {volume} {85}},\ \bibinfo
  {pages} {165126} (\bibinfo {year} {2012})}\BibitemShut {NoStop}%
\bibitem [{\citenamefont {Sinha}\ \emph {et~al.}(2025)\citenamefont {Sinha},
  \citenamefont {Pan},\ and\ \citenamefont {Bradlyn}}]{Sinha2025}%
  \BibitemOpen
  \bibfield  {author} {\bibinfo {author} {\bibfnamefont {S.}~\bibnamefont
  {Sinha}}, \bibinfo {author} {\bibfnamefont {D.~Y.}\ \bibnamefont {Pan}},\
  and\ \bibinfo {author} {\bibfnamefont {B.}~\bibnamefont {Bradlyn}},\
  }\bibfield  {title} {\bibinfo {title} {Computing the {${\mathbb{Z}}_{2}$}
  invariant in two-dimensional strongly correlated systems},\ }\href
  {https://doi.org/10.1103/PhysRevB.111.085116} {\bibfield  {journal} {\bibinfo
   {journal} {Phys. Rev. B}\ }\textbf {\bibinfo {volume} {111}},\ \bibinfo
  {pages} {085116} (\bibinfo {year} {2025})}\BibitemShut {NoStop}%
\bibitem [{\citenamefont {Bena}\ and\ \citenamefont
  {Montambaux}(2009)}]{Bena2009}%
  \BibitemOpen
  \bibfield  {author} {\bibinfo {author} {\bibfnamefont {C.}~\bibnamefont
  {Bena}}\ and\ \bibinfo {author} {\bibfnamefont {G.}~\bibnamefont
  {Montambaux}},\ }\bibfield  {title} {\bibinfo {title} {Remarks on the
  tight-binding model of graphene},\ }\href
  {https://doi.org/10.1088/1367-2630/11/9/095003} {\bibfield  {journal}
  {\bibinfo  {journal} {New J. Phys.}\ }\textbf {\bibinfo {volume} {11}},\
  \bibinfo {pages} {095003} (\bibinfo {year} {2009})}\BibitemShut {NoStop}%
\end{thebibliography}%

\appendix

\section{Applying the chiral symmetry to the Kane-Mele HK model with an orbital Zeeman field}
\label{app:chiral}

The chiral symmetry operator is here $S=\sigma_y\otimes \tau_z$. We can diagonalize $S$ like \begin{equation}
    S'=U^\dagger SU = \begin{pmatrix}
		1&0&0&0\\
		0&1&0&0\\
		0&0&-1&0\\
		0&0&0&-1
	\end{pmatrix},
\end{equation} with the unitary matrix \begin{equation}
	 U = \frac1{\sqrt2}\begin{pmatrix}
	 	1&0&1&0\\
	 	0&1&0&1\\
	 	i&0&-i&0\\
	 	0&-i&0&i
	 \end{pmatrix}.\end{equation} Then, in this new basis, we have \begin{equation}
	     U^\dagger HU = \begin{pmatrix}
	         \mathbf{0} & C^\dagger \\ C & \mathbf{0}
	     \end{pmatrix},~ C = \begin{pmatrix}
	h-B\text{e}^{i\theta} & T^* \\ T & -h-B\text{e}^{-i\theta}
\end{pmatrix}.
	 \end{equation} For the vector $\begin{pmatrix}
	     u\\ v
	 \end{pmatrix}$ to be an eigenstate of $U^\dagger HU$, we have the condition \begin{equation}
	U^\dagger HU \begin{pmatrix}
		u \\ v
	\end{pmatrix} = E \begin{pmatrix}
		u \\ v
	\end{pmatrix} \Rightarrow \begin{cases}
		C^\dagger v = Eu \\ Cu = Ev
	\end{cases},
\end{equation} so that we only have to diagonalize the $2\times 2$ matrix $C^\dagger C$ because we have the two chiral solutions with opposite energies \begin{equation}
    \begin{cases}C^\dagger C u=E^2 u \\ \displaystyle v_\pm= \frac{C u}{\pm\sqrt{E^2}} \end{cases}.
\end{equation} Using this method, we find the eigenstates of the Hamiltonian for low energies around the Dirac points and we can compute the Berry curvature at these points.

\section{Non-integer Berry curvature integral} \label{app:int_berry}

We call gauge transformation the operation on state $\ket{u}\rightarrow \text{e}^{\text{i}\varphi(\mathbf{k})} \ket{u}$ with $\varphi(\mathbf{k})$ a phase.

\subsection{General proof of the quantization}\label{app:proof_integer}

First, we come back to the proof showing that the integral of the Berry curvature over the full Brillouin zone is an integer number times $2\pi$. We call $\cal B$ the Berry curvature and $\vec{\cal A}$ the Berry connection. If the model is topological, there exists points in the Brillouin zone where the eigenstate is not smooth. We call these points $A_i$ and choose the edge of the Brillouin zone such that none of these points is on it. For each point, we choose a contour $\partial C_i$ surrounding it but such that $A_i\notin \partial C_i$. The set of these $\partial C_i$ is noted $S_\text{not smooth}$. We complete the set of contours by contours $\partial C_a$ such that the sum of these contours is the full Brillouin zone. This second set is noted $S_\text{smooth}$. We define $S_C = S_\text{not smooth}\cup S_\text{smooth}$. Then, because the eigenstate is smooth along all of these paths, we can write the Berry phase $\delta \phi$ accumulated on a closed path $\partial C\in S_C$ as \begin{equation}\delta\phi_{\partial C} = \oint_{\partial C} \vec{\cal A}\cdot\mathbf{dk}.\end{equation}
Now, inside a closed path $\partial C$, a surface that we call $\partial S$, \textit{we assume that we can find a gauge where the eigenstate is smooth} (if $\partial C\in S_\text{smooth}$ the eigenstate is already smooth, if $\partial C\in S_\text{not smooth}$, it is not, so we have to change the gauge). In either case, the eigenstate is now smooth and we can use the Stokes theorem to obtain \begin{equation}\delta\phi_{\partial C} = \iint_{\partial S} \mathcal{B} ~\text{d}^2\mathbf{k}~(\text{mod}~2\pi)\end{equation} because $\delta\phi$ and $\mathcal{B}$ are both invariant under gauge transformation. We then sum $\delta\phi$ over the full Brillouin zone \begin{equation} \phi=\sum_{\partial C} \delta\phi_{\partial C} = \oint_{\cal C_{FBZ}} \vec{\cal A}\cdot\mathbf{dk}.\end{equation} Because the first Brillouin zone is periodic, we get $\phi=0$ so that \begin{equation}\iint_{FBZ} \mathcal{B}~\text{d}^2\mathbf{k} = 0~(\text{mod}~2\pi).\end{equation}

\subsection{Our situation}

In our specific case, in the situation with a band Zeeman field, we can write the Berry connection of the ground state $\alpha\ket{\mathbf{k}-\uparrow} + \beta\ket{\mathbf{k}-\downarrow}$ like \begin{equation}\vec{\cal A} = \abs{\alpha}^2\vec{\cal A}_\uparrow + \abs{\beta}^2 \vec{\cal A}_\downarrow,\end{equation} with $\vec{\cal A}_\sigma$ the Berry connection of the original Kane-Mele bands as it is the case for the Berry curvature (proof in Eq.~(\ref{eq:berry_computation})). For each part $\vec{\cal A}_\sigma$, this is a Haldane model, so we know there exists a gauge such that we can apply the Stokes theorem for the spin isolated two-band systems where we can use the invariance of these Berry phases under gauge transformations \begin{align}\delta\phi_\uparrow &= \oint_{\partial \cal C} \vec{\cal A}_\uparrow \cdot\mathbf{dk} = \iint_{\partial S} \mathcal{B}_\uparrow~\text{d}^2\mathbf{k}~(\text{mod}~2\pi)\\
\delta\phi_\downarrow &= \oint_{\partial \cal C} \vec{\cal A}_\downarrow \cdot\mathbf{dk} = \iint_{\partial S} \mathcal{B}_\downarrow~\text{d}^2\mathbf{k}~(\text{mod}~2\pi).
\end{align} Finally we can show that \begin{equation}\iint_{FBZ} \mathcal{B}_\uparrow~\text{d}^2\mathbf{k} = \iint_{FBZ} \mathcal{B}_\downarrow~\text{d}^2\mathbf{k}~(\text{mod}~2\pi) = 0~(\text{mod}~2\pi).\end{equation} We write these two values $2\pi C_\sigma$ and we have an integral which is not necessary an integer \begin{equation}\frac1{2\pi} \iint_{FBZ} \mathcal{B}~\text{d}^2\mathbf{k} = \abs{\alpha}^2 C_\uparrow + \abs{\beta}^2 C_\downarrow.\end{equation}
The key observation is that $\phi = \oint_{FBZ} \vec{\cal A} \cdot\mathbf{dk}$ is zero because of the periodicity of the Brillouin zone but that \begin{equation}\phi\neq \iint_{FBZ} \mathcal{B}~\text{d}^2\mathbf{k}~(\text{mod}~2\pi)\end{equation} because the Stokes theorem does not apply. Indeed, we can show that there always is at least one point ($\pm\mathbf{K}$) where we cannot find a gauge for which $\vec{\cal A}_\uparrow$ and $\vec{\cal A}_\downarrow$ are both smooth (and then the total $\vec{\cal A}$ cannot be smooth) so that we cannot apply the proof of App.~\ref{app:proof_integer} (the italicized assumption does not hold here). We show this last point in the next section.

\subsection{Why is there a point where you cannot find a proper gauge}

We focus on the Kane-Mele model with $\Delta=0$. Under a band HK interaction and a band Zeeman field, we can write the $\mathbf{k}$-dependent ground state like ($\theta$ and $\phi$ are independent of $\mathbf{k}$) \begin{equation}
	\ket{u_0(\mathbf{k})} = \cos\left(\frac\theta2\right) \ket{\mathbf{k}-\uparrow} + \text{e}^{\text{i}\psi}\sin\left(\frac\theta2\right)\ket{\mathbf{k}-\downarrow}.
\end{equation} Defining $E_\mathbf{k} = \sqrt{\abs{t\gamma_\mathbf{k}}^2+h_z(\mathbf{k})^2}$, we can explicitly write \begin{align}
	\ket{\mathbf{k}-\uparrow} &= \begin{pmatrix}
		\alpha_\mathbf{k} \\ \beta_\mathbf{k} \text{e}^{i \arg(t\gamma_\mathbf{k})} \\ 0 \\0
	\end{pmatrix}, \\
	 \ket{\mathbf{k}-\downarrow} &= \text{e}^{\text{i}\varphi(\mathbf{k})} \begin{pmatrix}
		0\\0 \\ \beta_\mathbf{k} \\ \alpha_\mathbf{k}\text{e}^{\text{i}\arg(t\gamma_\mathbf{k})}
	\end{pmatrix}, \\ 
	\alpha_\mathbf{k} &= \sqrt{\frac{E_\mathbf{k}-h_z(\mathbf{k})}{2E_\mathbf{k}}}, \\
	\beta_\mathbf{k} &= \sqrt{\frac{E_\mathbf{k}+h_z(\mathbf{k})}{2E_\mathbf{k}}},
\end{align} and $\varphi(\mathbf{k})$ is a function to be specified (we have total freedom about this choice because the Hamiltonian is block-diagonal in spin). Around the Dirac points $\xi\mathbf{K}$, we can write \begin{equation}
	h_z(\xi\mathbf{K}+\mathbf{q})\approx -\xi 3\sqrt3 t'\sin(\phi) \equiv h_\xi,
\end{equation} and because $t\gamma_{\xi\mathbf{K}+\mathbf{q}} \approx \hbar v(\xi q_x-\text{i} q_y)\equiv \gamma_\xi(\mathbf{q})$ we can approximate \begin{align}
	\alpha_{\xi\mathbf{K}+\mathbf{q}}&\approx \sqrt{\left(\frac{\abs{\gamma_{-\xi}(\mathbf{q})}}{2h_\xi}\right)^2 + \frac{1-\text{sgn}(h_\xi)}2}, \\
	 \beta_{\xi\mathbf{K}+\mathbf{q}}&\approx \sqrt{\left(\frac{\abs{\gamma_{\xi}(\mathbf{q})}}{2h_\xi}\right)^2 + \frac{1+\text{sgn}(h_\xi)}2}.
\end{align} Then, we know that in one valley, $h_\xi<0$ while in the other $h_{-\xi}>0$ (if $t'>0$, we have $h_+<0$ and $h_->0$). In the valley with $h_\xi <0$, we have $\alpha_{\xi\mathbf{K}+\mathbf{q}}\approx 1$ and $\beta_{\xi\mathbf{K}+\mathbf{q}}\approx \abs{\gamma_{\xi}(\mathbf{q})}/(2h_\xi)$ so that we can write \begin{align}
	\ket{(\mathbf{k}=\xi\mathbf{K}+\mathbf{q})-\uparrow} &\approx_{\mathbf{q}\approx\mathbf{0}} \begin{pmatrix}
		1 \\ \gamma_{\xi}(\mathbf{q}) / (2h_\xi) \\ 0 \\ 0
	\end{pmatrix}, \\
	\ket{(\mathbf{k}=\xi\mathbf{K}+\mathbf{q})-\downarrow} &\approx_{\mathbf{q}\approx\mathbf{0}} \text{e}^{\text{i}\varphi(\mathbf{k})}\begin{pmatrix}
		0 \\ 0 \\ \abs{\gamma_{\xi}(\mathbf{q})}/(2h_\xi) \\ \text{e}^{\text{i}\arg(\gamma_{\xi}(\mathbf{q}))}
	\end{pmatrix} \\
	&= \begin{pmatrix}
	0 \\ 0 \\ \gamma_\xi(\mathbf{q})^* / (2h_\xi) \\ 1
	\end{pmatrix},
\end{align} choosing conveniently $\varphi(\mathbf{k})=-\arg(\gamma_{\xi}(\mathbf{q}))$ so that the ground state wavefunction $\ket{u_0(\mathbf{k})}$ is explicitly smooth around this point. In the valley with $h_{-\xi}>0$, keeping the same gauge, we can also write \begin{align}
	\ket{(\mathbf{k}=-\xi\mathbf{K}+\mathbf{q})-\uparrow} &\approx_{\mathbf{q}\approx\mathbf{0}} \begin{pmatrix}
		\abs{\gamma_{-\xi}(\mathbf{q})} / (2h_{-\xi}) \\ \text{e}^{\text{i}\arg(\gamma_{-\xi}(\mathbf{q})} \\ 0 \\ 0
	\end{pmatrix}, \\
	 \ket{(\mathbf{k}=-\xi\mathbf{K}+\mathbf{q})-\downarrow} &\approx_{\mathbf{q}\approx\mathbf{0}} \begin{pmatrix}
	0 \\ 0 \\ \text{e}^{-\text{i}\arg(\gamma_{-\xi}(\mathbf{q})} \\ \abs{\gamma_{-\xi}(\mathbf{q})}/{(2h_{-\xi})}
	\end{pmatrix}.
\end{align} The ground state writes then \begin{widetext}\begin{equation}
	\ket{u_0(\mathbf{k}=-\xi\mathbf{K}+\mathbf{q})} \approx_{\mathbf{q}\approx\mathbf{0}} \cos\left(\frac\theta2\right) \begin{pmatrix}
		\abs{\gamma_{-\xi}(\mathbf{q})} / (2h_{-\xi}) \\ \text{e}^{\text{i}\arg(\gamma_{-\xi}(\mathbf{q}))} \\ 0 \\ 0
	\end{pmatrix} + \text{e}^{\text{i}\psi} \sin\left(\frac\theta2\right) \begin{pmatrix}
	0 \\ 0 \\ \text{e}^{-\text{i}\arg(\gamma_{-\xi}(\mathbf{q}))} \\ \abs{\gamma_{-\xi}(\mathbf{q})}/{(2h_{-\xi})}
	\end{pmatrix}.
\end{equation}\end{widetext} This is not smooth. We suppose now that there exists a gauge such that $\ket{u_0(-\xi\mathbf{K}+\mathbf{q})}=\text{e}^{\text{i}\Theta(\mathbf{k})}\ket{u_0(-\xi\mathbf{K}+\mathbf{q})}$ is now smooth. To make the first term smooth, we have to find $\Theta(\mathbf{k})$ such that $\text{e}^{\text{i}\Theta(\mathbf{k})} \text{e}^{\text{i}\arg(\gamma_{-\xi}(\mathbf{q}))} = f(\mathbf{k})$ with $f$ a smooth function (obviously different from zero) around $-\xi\mathbf{K}$. We obtain immediately $\text{e}^{\text{i}\Theta(\mathbf{k})} = f(\mathbf{k}) \text{e}^{-\text{i}\arg(\gamma_{-\xi}(\mathbf{q}))}$. We can proceed similarly with the second term and the second condition writes $\text{e}^{\text{i}\Theta(\mathbf{k})} = g(\mathbf{k}) \text{e}^{\text{i}\arg(\gamma_{-\xi}(\mathbf{q}))}$ with $g$ smooth around $-\xi\mathbf{K}$. The two conditions gives then \begin{equation}
	\frac{g(\mathbf{k})}{f(\mathbf{k})} = \text{e}^{-2\text{i}\arg(\gamma_{-\xi}(\mathbf{q}))}.
\end{equation} The term on the left-hand side is smooth ($f$ is different from zero) but the term on the right-hand side is not smooth for $\mathbf{k}=-\xi\mathbf{K}$. Hence, such a gauge $\Theta(\mathbf{k})$ does not exist.

\end{document}